\documentclass[aps, nofootinbib, twocolumn,groupedaddress,floatfix,nofootnotesinbib,preprintnumbers]{revtex4}
\usepackage{graphicx}
\usepackage{epsfig}
\usepackage{epstopdf}
\usepackage{rotating}
\usepackage[mathscr]{eucal}
\usepackage{amsmath,amssymb,bm,graphics}
\usepackage[letterpaper=true]{hyperref}

\let\mathbf=\bm

\newif\ifarxiv
\arxivfalse

\begin{document}
\def\Nfour{\mathcal N\,{=}\,4}
\def\Ntwo{\mathcal N\,{=}\,2}
\def\Nc{N_{\rm c}}
\def\Nf{N_{\rm f}}
\def\x{\mathbf x}
\def\q{\mathbf q}
\def\f{\mathbf f}
\def\v{\mathbf v}
\def\C{\mathcal C}
\def\w{\omega}
\def\vs{v_{\rm s}}
\def\S{\mathcal S}
\def\half{{\textstyle \frac 12}}
\def\twothirds{{\textstyle \frac 23}}
\def\third{{\textstyle \frac 13}}
\def\t{\mathbf{t}}
\def\T{\mathcal {T}}
\def\O{\mathcal{O}}
\def\E{\mathcal{E}}
\def\p{\mathcal{P}}
\def\H{\mathcal{H}}
\def\uh{u_h}
\def\R{\ell}
\def\Ro{\chi}
\def\del{\nabla}
\def\eps{\hat \epsilon}
\def\nn{\nonumber}
\def\K{\mathcal K}
\def\inf{\epsilon}
\def\cs{c_{\rm s}}
\def\A{\mathcal{A}}
\def\e{{e}}
\def\r{{\xi}}
\def\x{{\mathbf x}}
\def\w{{w}}
\def\rr{{\xi}}
\def\uo{{u_*}}
\def\u{{\mathcal U}}
\def\G{\mathcal{G}}
\def\Deltax{\Delta x_{\rm max}}
\def\L{{\bm L}}

\title
{Black Topologies Production in Extra Dimensions}

\author{Anastasios~Taliotis\footnotemark}

\affiliation
   {Department of Physics, University of Crete, Greece}

\date{March 29, 2012}

\begin{abstract}
The configuration resulting after a collision of gravitational sources in a higher dimensional space with extra dimensions is investigated. Evidence is found that as the energy increases, there is a phase transition in the topology of the black object that is being formed: from the Black Hole to the Black String topology. An intuitive mechanism for the way the transition takes place is being proposed. The transition occurs at a finite value of the energy where an upper and a lower bound is found. Furthermore, at low energies the compact dimension behaves as an extended one while at high energies the extra dimension seems to decouple. Finally, the implications about the Gregory-Laflamme instability, the implications to the accelerators as well as holographic implications are being discussed.
\end{abstract}

\preprint{CCTP}


\maketitle
\iftrue
\def\thefootnote{\fnsymbol{footnote}}
\footnotetext[1]{Email: \tt taliotis@physics.uoc.gr}
\def\thefootnote{\arabic{footnote}}
\fi

\parskip	2pt plus 1pt minus 1pt

\noindent \section{{Context}}  

We analyze the problem of a black object formation in the presence of compact-extra dimensions in flat backgrounds when shock-waves collide. In particular, we attempt to follow the evolution of the topology \cite{Myers:1986un,Myers:2011yc,Emparan:2001wn,Kol:2002xz,Kol:2004ww,Asnin:2006ip,Kol:2003ja,Kol:2003if,Gorbonos:2005px,Emparan:2007wm,Emparan:2011ve} of the resulting (black) object as the energy involved in the collision, changes. We also estimate the entropy associated with the corresponding trapped surface in the spirit of \cite{Eardley:2002re,Giddings:2004xy,Kiritsis:2011yn,Aref'eva:2009ng,Aref'eva:2009wz,Aref'eva:2009kw,Nastase:2008hw,Lin:2010cb,Kovchegov:2009du,Gubser:2009sx,Gubser:2008pc,DuenasVidal:2012sa,DuenasVidal:2010vi,AlvarezGaume:2008fx}. For concreteness we consider five extended and one compact dimensions where an analytic shock wave solution in closed form is possible. The literature is rich regarding the investigation of shock-waves in gravity. For instance, in flat spaces they have been studied from  \cite{D'Eath:1992hb,D'Eath:1992hd,D'Eath:1992qu,Aichelburg:1970dh,Dray:1984ha,Hawking:1969sw,Sfetsos:1994xa,Taliotis:2010az,Eardley:2002re,Constantinou:2011ju,Gal'tsov:2010me,Gal'tsov:2009zi,Herdeiro:2011ck} while for more generalized backgrounds they have been studied in \cite{Aref'eva:2009kw,Aref'eva:2009wz,Aref'eva:2012ar,Kang:2004jd,Giddings:2002cd,Lin:2010cb,Kovchegov:2007pq,Spillane:2011yf,Sfetsos:1994xa,Hotta:1992qy,Albacete:2008vs,Khlebnikov:2010yt,Khlebnikov:2011ka,Albacete:2009ji,Wu:2011yd,Albacete:2008ze, Taliotis:2009ne,Grumiller:2008va,Taliotis:2010pi}.

In this work it is argued that the there exists a finite energy of the shock where the trapped surface and as a consequence the produced Black Hole (BH) that will result immediately after, occupies all of the compact space. At that energy, a phase transition should appear: the localized apparent horizons should cover the compact dimension and merge yielding to a Black-String (BS) transition; that is the system seems to yield to a BH$\rightarrow$BS phase transition \cite{Kol:2002xz,Kol:2004ww,Asnin:2006ip,Kol:2003ja,Kol:2003if,Gorbonos:2005px,Harmark:2002tr,Harmark:2003eg,Karasik:2004ds,Chu:2006ce,Wiseman:2002zc,Sorkin:2004qq,Kleihaus:2006ee,Emparan:2007wm,Emparan:2011ve}. The energy $E$ where this occurs, in the 5+1 (compact) dimensions that we work, should be found numerically. The analytical investigation of this work, provides the bound
\begin{align}\label{etr}
 0.8 \lesssim  \mu_{mer}\equiv \frac{16 \pi G_6}{L^3}E_{mer}\lesssim29=\mbox{finite,}  \hspace{0.2in}L=2\pi R 
\end{align}
where $G_6$ and $R$ are the Newton constants in six dimensions and the radius of the compact dimension respectively. The dimensionless parameter $\mu_{mer}$ corresponds to the merging energy $E_{mer}$ where the topology transition takes place: the BH covers the whole compact dimension and becomes an (initially non-uniform) BS. The bound of (\ref{etr}) respects the numerical works of \cite{Sorkin:2003ka,Kudoh:2003ki,Kudoh:2004hs} for five space-time plus one compact dimensions. A detailed and pedagogical review on these topics is the work \cite{Harmark:2007md} which provides additional references while reference \cite{Chowdhury:2006qn}  examines the problem from a microscopical point of view (see also the related reviews \cite{Mathur:2005zp,Mathur:2005ai}).

\section{Setting up the problem}
Trapped surfaces are created when two shocks collide. The one shock moves along $x^-$ and we call it $\phi_{+}$ to distinguish it from the second that moves along $x^+$ and which we call  $\phi_{-}$. In our case the shocks will be taken identical and hence the subscripts will be soon dropped \footnote{A more complete set of notes on the theory of trapped surfaces may be found in the appendices of \cite{Kiritsis:2011yn}.}. In terms of metrics before the collision, one then has

\begin{align}\label{gmn}
ds^2&= -2dx^+dx^-+ dx^idx^i+(R d\theta)^2   \notag \\ &
 + \{ \phi_+(R\theta,r)\delta(x^+)(dx^+)^2  +(+\leftrightarrow-) \}
 , \hspace{0.1in}x_{\pm}<0
 \end{align}
where $i=1,2,3$, $r=\sqrt{(x^i)^2}$, $x^{\pm}=(x^0\pm x^4)/\sqrt{2}$, $\theta \in [-\pi,\pi]$ and 
\begin{align}\label{fi}
\phi_+&=\frac{2^{3/2} E G_6}{\pi R}\frac{1}{r} \left( 1+\frac{1}{e^{\frac{r}{R}+i \theta}-1} +\frac{1}{e^{\frac{r}{R}-i \theta}-1}  \right) \notag\\&
=\frac{2^{3/2} E G_6}{\pi R}\frac{1}{r} \frac{\sinh(r)}{\cosh(r)-\cos(\theta)}
\end{align}
A few remarks on $\phi_+$ of (\ref{fi}) follow:

\begin{itemize}

\item{ It is periodic with respect to the compact dimension $\theta$ as should and has $\theta \leftrightarrow -\theta$ symmetry.}

\item{It satisfies $\nabla_{\perp}^2\phi_+\equiv(\nabla_i^2+1/R^2\partial^2_{\theta})\phi_+=-16\sqrt{2}\pi G_6 E/R \delta^{(3)}(${\bf r}$)\delta(x^+)\delta(\theta)$ \footnote{In order to verify it one has to note that the fraction that involves the trigonometric quantities, in the limit where $r\rightarrow 0$ results to $2\pi \delta(\theta)$.}. Taking into account that $R_{++}=-\frac{1}{2}\nabla_{\perp}^2\phi_+=8\pi G_6 T_{++}$ it is deduced that the shock $\phi_{+}$ is associated with a point-like stress-tensor $T_{++}=E\sqrt{2}/R \delta^{(3)}(${\bf r}$)\delta(x^+)\delta(\theta)$ which moves with the speed of light along  $-x^4$ axis.}

\item{It happens that $\phi_+$ satisfies the Poisson equation of gravity in 3 (non compact) + 1 (compact) dimensions. At large distances $r\gg R$ it behaves as $\sim 1/r$, that is as in Newtons law in 3D. On the other hand, al small distances $r\ll R$ and assuming that the measurement of the gravitational potential is restricted on some brane with $\theta=0$ in the spirit of \cite {Randall:1999ee}, then (\ref{fi}) behaves as $\sim R/r^2$. This behavior, coincides with the ideas of \cite{Antoniadis:1998ig,ArkaniHamed:1998rs,ArkaniHamed:1998nn}. However, $\phi_+$ here will be used in a different context.}

\item{The ordering of limits $\lim_{r \to 0}$ and $\lim_{\theta \to 0}$ does not commute.}

\end{itemize}

Associated with the shock-wave $\phi_{\pm}$ in (\ref{gmn}),  we parametrize (half of the) trapped surface $S_{\pm}$ by
\begin{align}\label{TS}
x^{\pm}=0  \hspace{0,2in} x^{\mp}+\frac{1}{2}\psi_{\pm}(\theta,r)=0
\end{align}
where $\psi_+$ remains to be determined.
The function $\psi_+$ satisfies the following differential equation
\begin{align}\label{de}
 \nabla^2_{\perp} (\psi_{\pm}-\phi_{\pm})=0.
\end{align}

It is pointed out that $(\nabla^2_{\perp} ) \phi_{\pm}$ provides a source term for  $(\nabla^2_{\perp} ) \psi_{\pm}$. The missing ingredient is the boundary conditions and are given by
 \begin{align}\label{BC}
\psi_{\pm} \Big |_C=0 \hspace{0.4in}  \sum_{i=R\theta,r}\left[ \nabla_i \psi_+ \nabla_i \psi_- \right]  \Big |_{C}=8
 \end{align}
 for some curve $C$ which defines the boundary of the trapped surface and where both, $S_+$ and $S_-$ end. The produced entropy is then bounded below by the area of the surface obtained by adjoining the two pieces of the trapped surface associated with each of the shocks. It is given by
 \begin{align}\label{ent1}
S_{prod} \geq S_{trap}=2\times \frac{R}{4G_6} \int_C d\theta d^3 {\bf x} = \frac{2 \pi R}{3G_6} \int_{\theta_1}^{\theta_2}  r^3(\theta,E)d\theta
\end{align}
where the (generalized) curve $C$ defines the boundary of the trapped surface $S_+$ and $S_-$ which are identical; thus the overall factor of 2. The integral with respect to the non compact direction gives $r^3$ when considering a head-on collision. Typically, $\theta_1$, $\theta_2$ and $r$ carry the information of the shock $\phi$ \footnote{We drop the subscripts $+,-$ from $\phi$'s and $\psi$'s from now on.}.

The boundary conditions define a curve $C$
\begin{align}\label{C}
r(\theta,E)
\end{align}
which is explicitly given below, in equation (\ref{bdry}) in two extreme limits. Evidently, the dimensionless quantity $EG_6/R^3$ sets the high and the low energy limit of the process. The last step would be to (numerically) solve equation (\ref{C}) for $r$ and integrate for several (but fixed) values of the parameter 
\begin{align}\label{x}
x \equiv R^3 \pi / EG_6. 
\end{align}
In what follows, we will derive analytical results for the two extreme cases of high and low energies.

\section{Analysis and Results}
\begin{figure}
\includegraphics[scale=0.62]{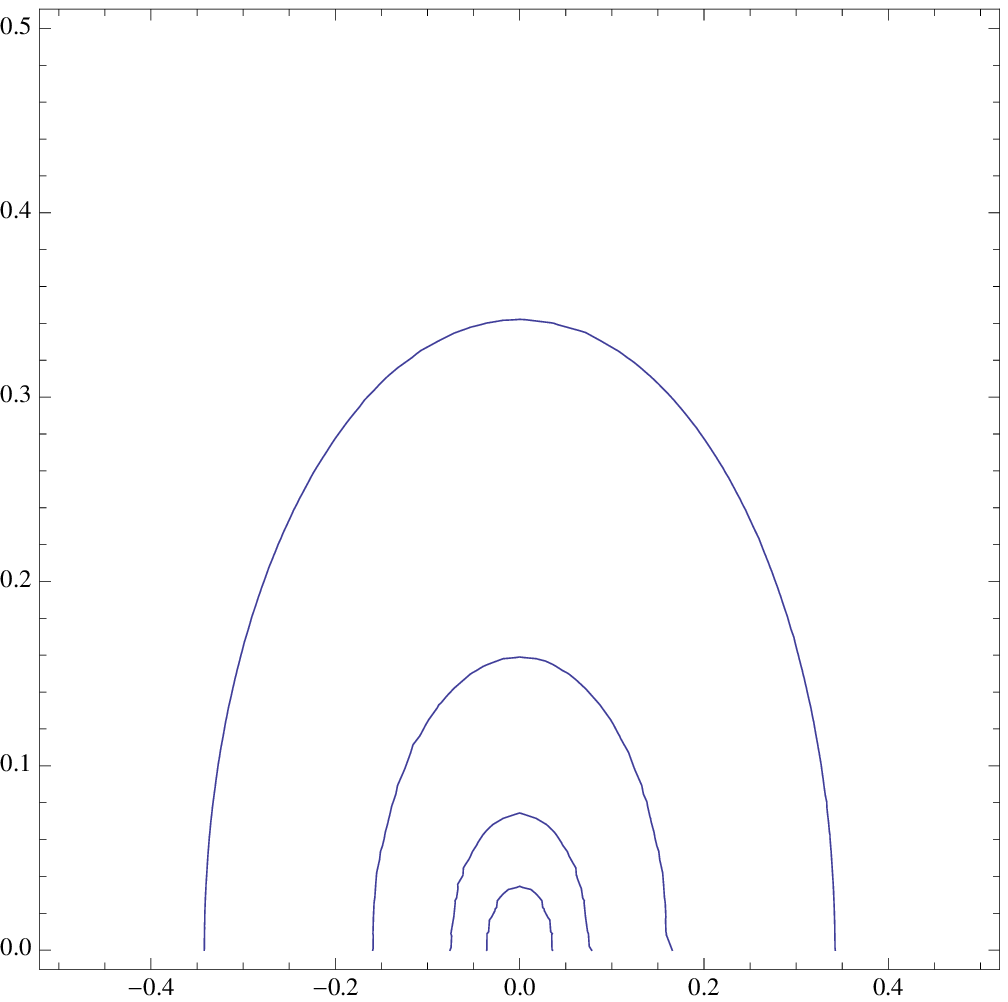}
\includegraphics[scale=0.62]{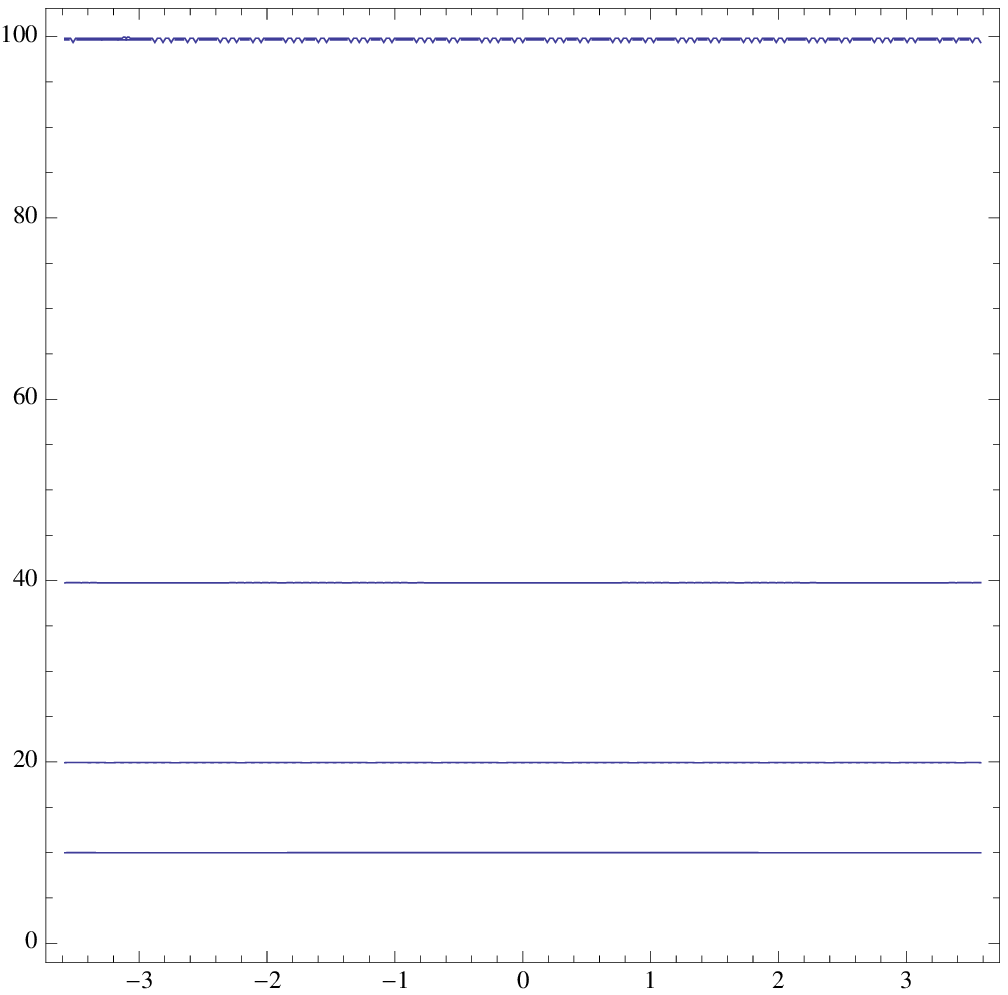}
\caption
    {\label{LHE}
     Plot of the boundary $C$ of the trapped surface in the two extreme limits as found from analytical considerations. Horizontal axis: $\theta$, vertical axis: $u=r/R$ (the same applies for the plots of the figures that follow). Top plot refers to low energies and yields to a $D^4$ topology (BH). As $x$ increases in the range $x=10^2,10^3,10^3,10^5$ the circles become smaller). Bottom plot refers to high energies (BS) with topology $D^3\times S^1$. As $x$ decreases in the range $x=10^{-2},10^{-2.6},10^{-3.2},10^{-4},$ the lines move upwards.
  }
\end{figure} 

\subsection{BH $\rightarrow$ BS Phase Transition}
In this section it is argued that the resulting system covers the whole compact dimension at a finite energy and passes from a BH phase to a BS phase. This is done by analyzing the low and the high energy asymptotics. In the low energy asymptotics we find a trapped horizon with topology\footnote{In what follows, we use the symbol $D^4$ (and $D^3$) for the 4-disk (and 3-disk) imbedded in (ignoring time) $\mathbb{R}^{4+1}$ where plus one refers to the compact dimension. We prefer the symbol $D^n$ instead of the more traditional $B^n$ (unit ball) in order to highlight that these disks are imbedded in a higher dimensional space. For instance $D^4$ can be thought as the intersection of a 5-ball with the $x^1,x^2,x^3$ hyperplane.} $D^4$. 
After the horizon will obtain rapidity dependence\footnote{ That is after the horizon will expand along $x^4$; the direction where the shocks are moving initially.}, it will presumably evolve to an $ S^4$ horizon (likely to a Schwarzschild BH). At high energies on the other hand, we find $D^3\times S^1$ to be evolved to $S^3\times S^1$; hence the BH $\rightarrow$ BS phase transition. These limiting geometries that we find below after the shock-waves collide, agree with the expected behavior in the literature \cite{Kol:2002xz}.

The starting point is the general solution to (\ref{de}) which has the form
\begin{align}\label{sol}
\psi=\phi+\phi_h, \hspace{0.1in}\mbox{where}  \hspace{0.1in} \nabla_{\perp}^2\phi_h=0.
\end{align}
and where all the three functions generally depend on the energy $E$. For the two extreme cases that we are interested, the solutions $\phi_h$ of the homogeneous degenerate to two (different) constants respectively. In this case, the second (non-linear) boundary condition of (\ref{BC}), becomes
\begin{align}\label{bdry}
& (-2u + 2u \cos(\theta)\cosh(u)-2 \cos(\theta)\sinh(u)+\sinh^2(2u))^2\notag\\ &
-4 x^2 u^4(\cos(\theta)-\cosh(u))^4 \notag\\ &
+4 u^2\sin^2(\theta)\sinh^2(u)=0,\hspace{0.1in}u=r/R.
\end{align}
\begin{figure}
\includegraphics[scale=0.6]{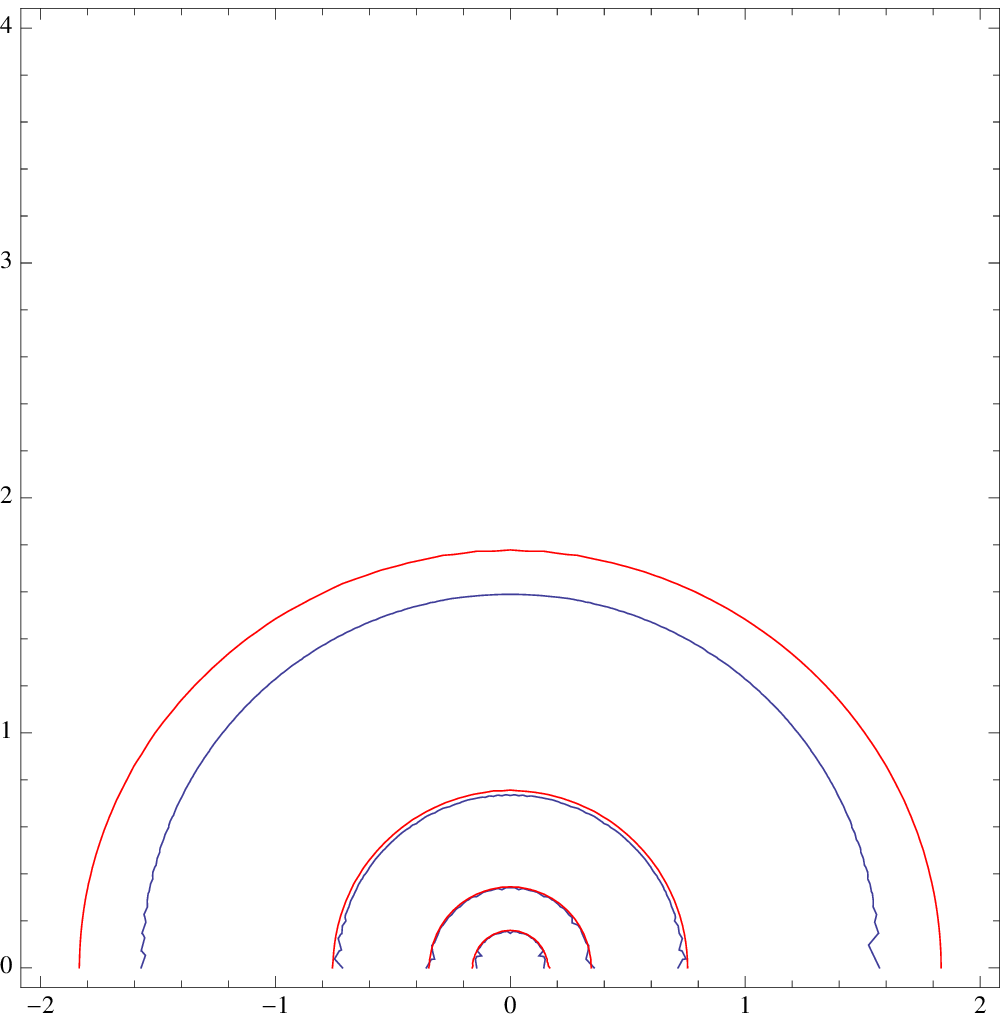}
\includegraphics[scale=0.6]{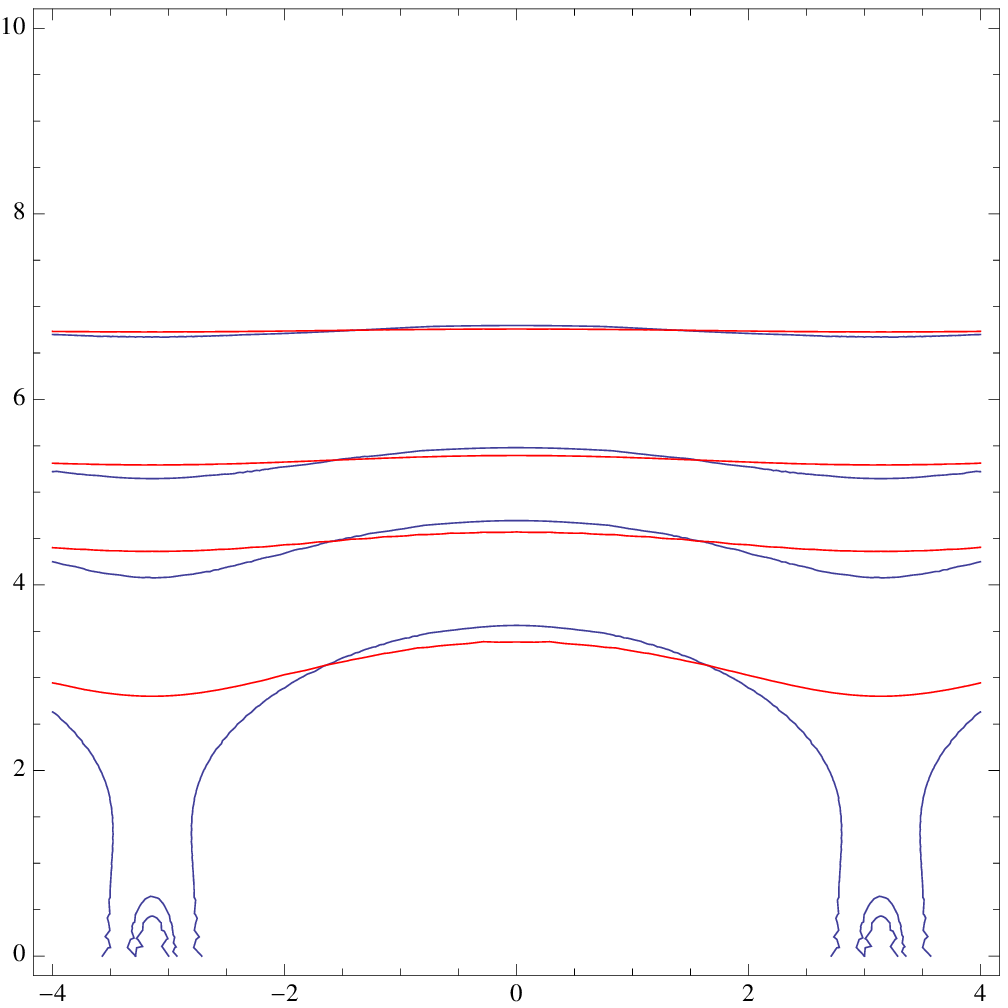}
\caption
  {\label{merge}
  Trapped boundary for the two distinct boundary conditions (\protect{\ref{BC}}); red and blue curves correspond to the first and second boundary condition respectively. Top panel: $\phi_h^{x\gg1}=C^{x\gg 1}=-2^{7/6} R x^{-1/3}$ and $x=1,8,10^2,10^3$. The pairs of curves shrink as $x$ increases and soon they merge (third pair of curves for $x=8$; low energy limit that results to a BH). Bottom panel: $\phi_h^{x\ll1}=C^{x\ll 1}=-2^{3/2} R x^{-1/2}$ and $x=0.1$, $0.035$ (dashed plots), $0.02$. The pairs of curves move upwards as $x$ decreases and soon they merge (top pair of curves with $x=0.02$; high energy limit that results to a BS).} 
\end{figure} 
\subsection{Limiting Behaviours}

I. Low Energy Asymptotics: $x\gg1$.
\vspace{0.05in}

The claim is that the choice $\phi_h\equiv C_x=-2^{7/6} R x^{-1/3}$ and the boundary $C: u^2+\theta^2=2^{4/3}x^{-2/3}$ satisfy both of the conditions of (\ref{BC}). Indeed, the second boundary condition for large $x$ yields $C: x^2(u^2+\theta^2)^3=16$ while the first one yields $u^2+\theta^2=-2^{5/2}R/xC_x$ from where one identifies that $\phi_h\equiv C_x=-2^{7/6} R x^{-1/3}$. This configuration (BH) is shown in the top plot of fig. \ref{LHE} and \ref{merge}.

 The (lower) entropy bound may now be found. Using (\ref{ent1}) immediately yields
\begin{align}\label{entLE}
S_{trap} & =  \frac{2 \pi R^4}{3G_6} 2\int_{0}^{2^{2/3}x^{-1/3}} (2^{4/3}x^{ - 2/3}-\theta^2)^{3/2}d\theta  \notag\\&
=2^{2/3} \pi ^2 \frac{ R^4}{G_6}  x^{-4/3}=(2 \pi)^{2/3}(E^4G_6)^{1/3}.
\end{align}
There are two related observations about this result. The first one is that the $S_{prod}|_{x \gg1}$ is independent from the radius of the compact dimension $R$. The second observation is that the result agrees exactly with the result of Giddings and Eardley \cite{Eardley:2002re} for $D=6$ extended space-time dimensions and which states that in $D$ dimensions the trapped entropy is
\begin{align}\label{GE}
S_{prod}^D \sim \frac{1}{G_D} \left(\frac{EG_D}{\Omega_D}\right)^{\frac{D-2}{D-3}}
\end{align}
with $\Omega_D$ the solid angle. The physics behind this result is that at low energies, the dynamics are not capable to resolve the finiteness of the compact dimension. Therefore, the extra dimension behaves as being (also) extended. In other words, the resulting BH has size much smaller than $R$. This statement, according to (\ref{x}), is consistent with $x\gg1$ and the fact that the final BH will have a small (presumably) Schwarzschild  horizon $r_h \sim G_6E/R^2 \ll R$ (see \cite{Kol:2002xz}).

\vspace{0.15in}
II. High Energy Asymptotics: $x\ll1$.
\vspace{0.05in}

The claim is that the choice $\phi_h\equiv C_x=-2^{3/2} R x^{-1/2}$ and the boundary $C: u=x^{-1/2}$ satisfy both of the conditions of (\ref{BC}). Indeed, the second boundary condition for large $x$ and yields $C: x^2 u^4=1$ while the first one $2^{3/2}x^{-1} R u+C_x=0$ from where one identifies that $\phi_h\equiv C_x=-2^{3/2} R x^{-1/2}$. The trapped boundary is evidently independent on $\theta$ and yields to the BS configuration. This is depicted on the bottom plot of fig. \ref{LHE} and \ref{merge} for several large values of $E$. The $S_{prod}$ is then trivial to find and yields
\begin{align}\label{entHE}
&S_{trap}= \frac{2\pi R^3}{3G_6} \left(x^{-1/2} \right)^3  \times R\int_{-\pi}^{\pi}\theta= \notag\\ &
 \frac{2\pi}{3}  \frac{R^3}{G_6} x^{-3/2} L = \frac{8}{3\sqrt{2}} \pi  (E^3G_5)^{1/2}, \hspace{0.1in} G_5\equiv \frac{G_6}{L} . 
\end{align}
The result agrees with the result of \cite{Eardley:2002re}, equation (\ref{GE}), with $D=5$ and an effective Newton's constant $G_5=G_6/L$ and also the expectations of \cite{Kol:2002xz,Giddings:2001bu}\footnote{This work deals with 4 extended and one compact dimensions.}.

The physics behind this result becomes clear once we make the logical hypothesis that the final state will be a Schwarzschild object in the sense that its size ($r_h$) will increase with the energy and occupy the whole compact dimension. In fact, for high enough energies, it will be true that $r_h\gg R$ and hence $R$ may be neglected and can appear in the expression of the entropy only trivially. Indeed the compact dimension appears as a product space; this is the BS solution. The only trace of the compact dimension is in the effective Newton's constant which changes from $G_6$ to $G_5\equiv \frac{G_6}{2 \pi R}$.

\section{Possible Paths: BH$\rightarrow$BS}
The solution $\phi_h=C_x=$constant (see (\ref{de})) yields to two generally distinct families (that are indexed by $x$) of trapped boundaries. The one family is given by the curves defined by the first and the other by the second boundary condition of (\ref{BC}). The two families coincide at large $x$ (small $E$; see top plots of fig. \ref{merge}) and also at small $x$ (large $E$; see bottom plots of fig. \ref{merge}); only in these two cases the trapped surfaces of fig. \ref{LHE} are the desired solutions to the boundary value problem defined by equations (\ref{de}) and (\ref{BC}). 
\begin{figure}
\includegraphics[scale=0.53]{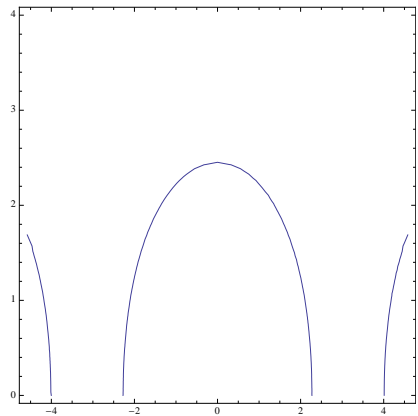}
\caption
    {\label{rest} As the energy increases, it is expected that the``nearby" BHs will grow and approach each other.}
\end{figure} 

\begin{figure}[t]
\includegraphics[scale=0.47]{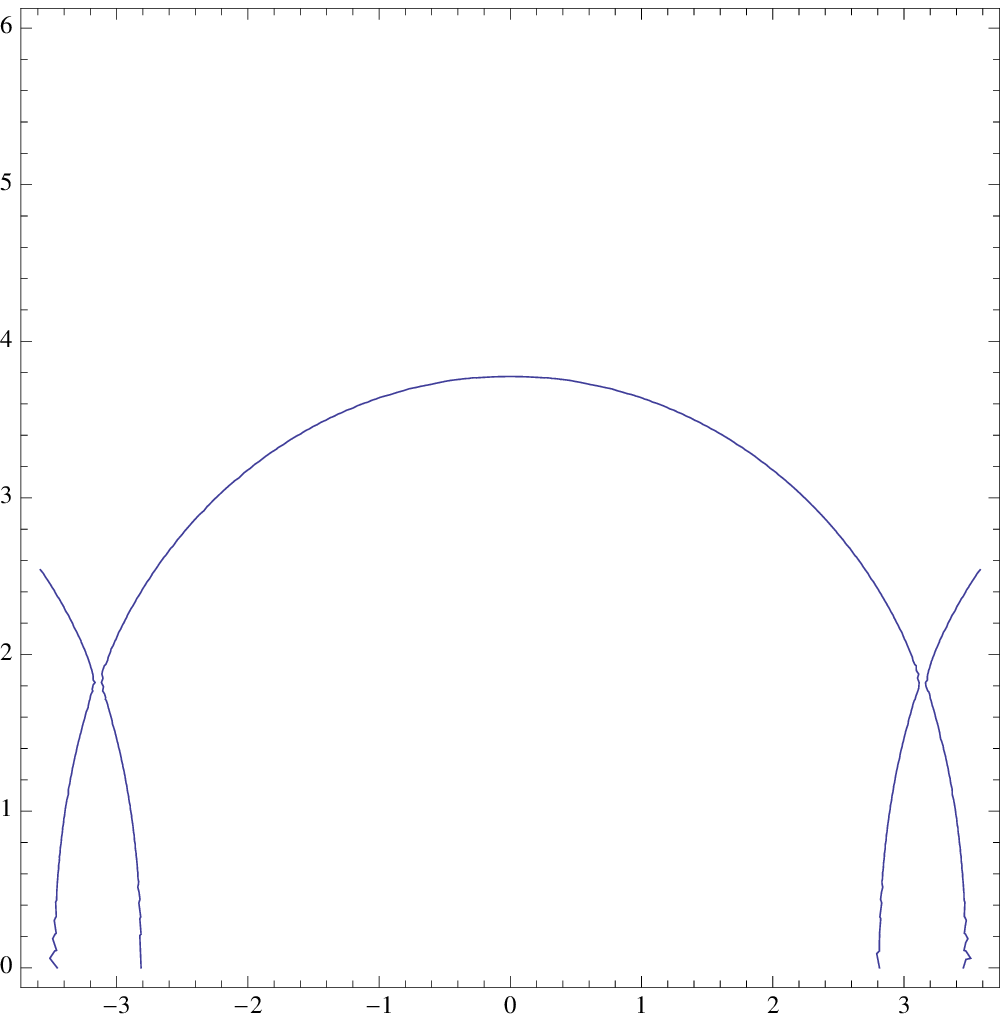}
\includegraphics[scale=0.47]{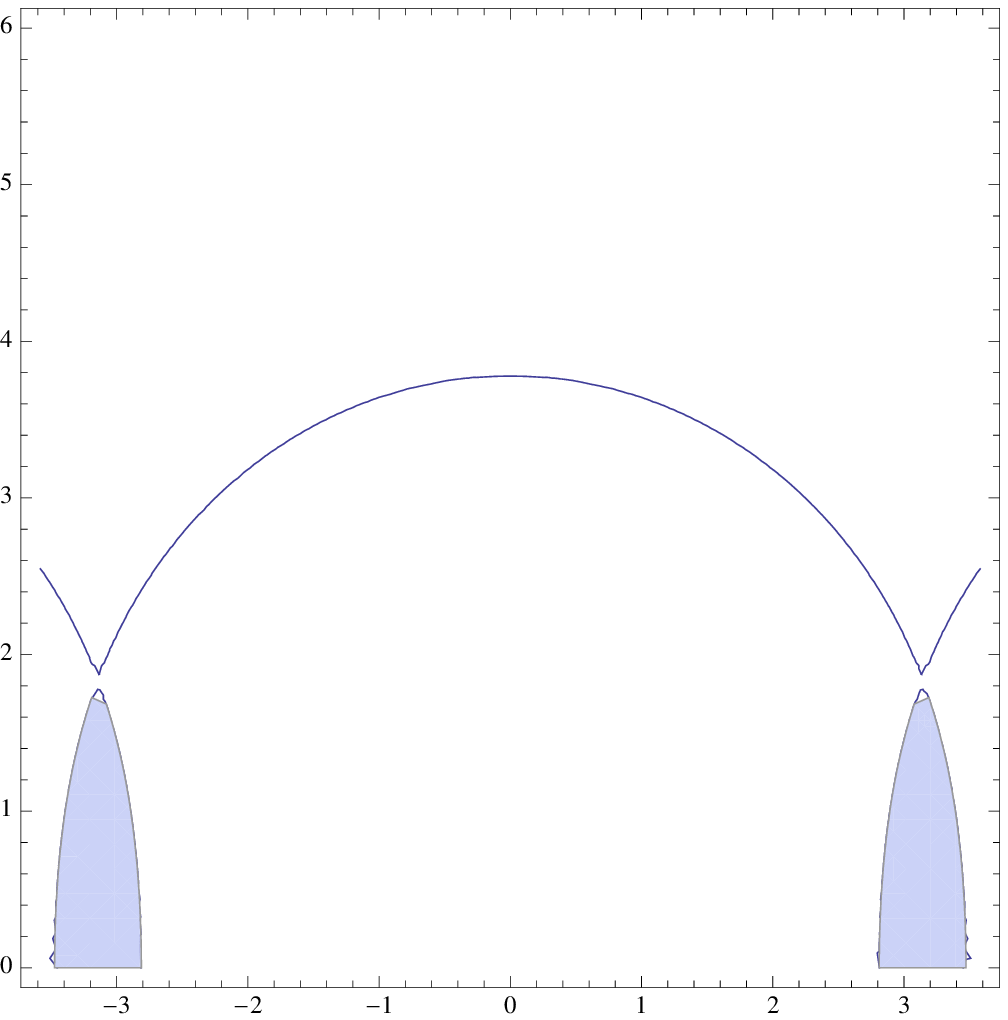}
\includegraphics[scale=0.47]{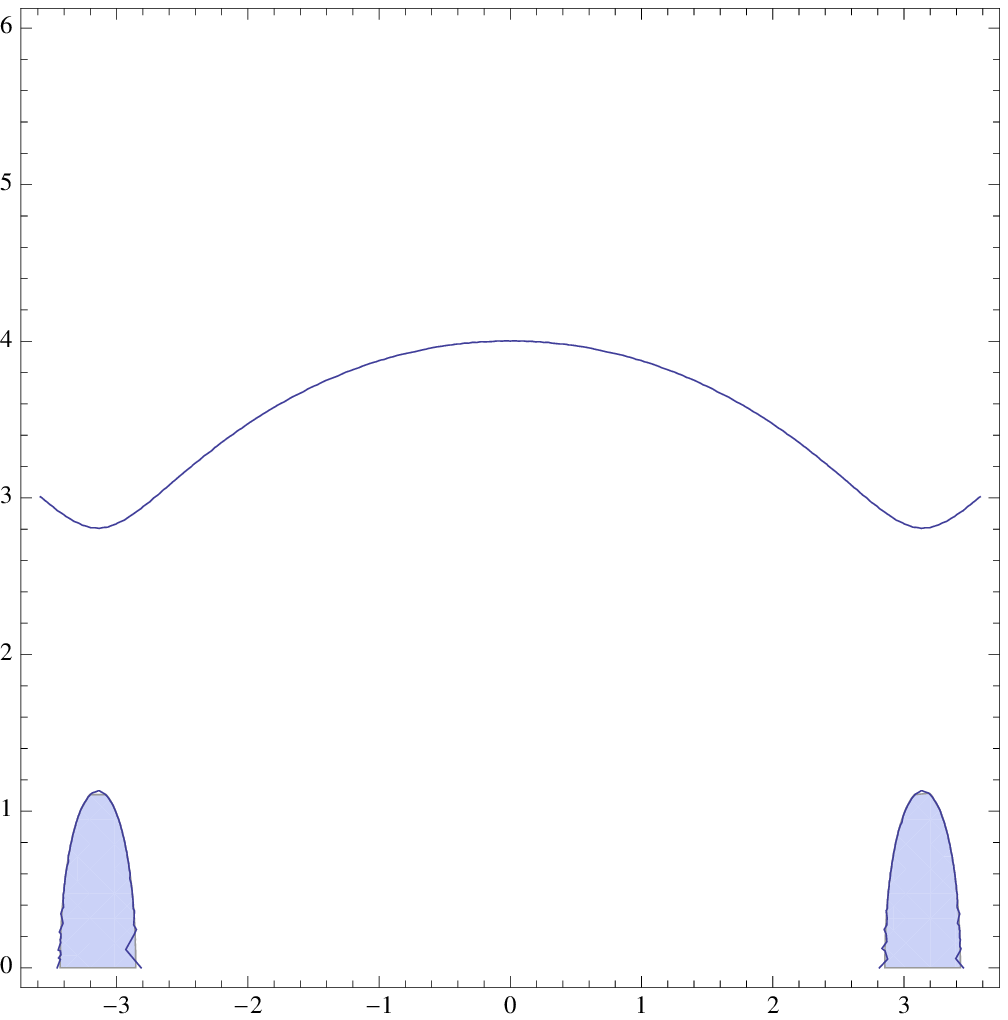}
\caption
  {\label{Path1}
  The curves result from equation (\protect{\ref{bdry}}).  The value of $x$ decreases from the top to the bottom plot. In the top plot, the trapped horizon which is not convex, covers the whole compact dimension because it extends up to values of $\theta=\pm \pi$. Increasing the energy further (two bottom plots), seems to yield to a phase transition: the resulting BH begins to wind the compact dimension changing the topology from $D^4$ to $D^3 \times S^1$. This is  a non-uniform BS topology. As the energy increases further ($x$ decreases), the curves move upwards ending up to straight lines independent on $\theta$ and matching those of fig. \protect{\ref{LHE}} (uniform BS configuration). It is pointed out that the shaded areas consisting of the two (half) ellipses left and right of $r=0$ are not a part of the trapped surface. There presence is to shield the image charges (see sec. \protect{\ref{ES}} ).} 
\end{figure}

\begin{figure}[t]
\includegraphics[scale=0.6]{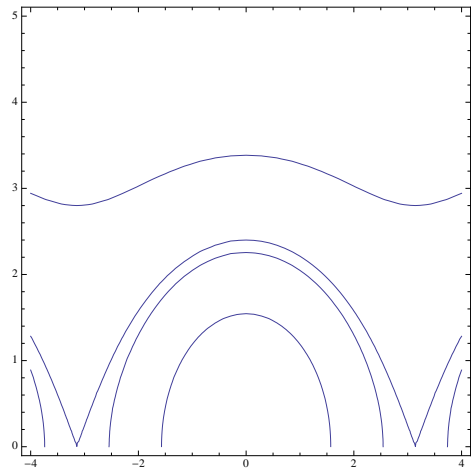}
\caption
  {\label{Path2}[t]
  Trapped boundary resulting from first condition of (\protect{\ref{BC}}) with  $\phi_h^{x\ll1}=C^{x\ll 1}=-2^{3/2} R x^{-1/2}$. As energy increases ($x$ decreases), the curves move upwards ending up to straight lines and matching those of fig. \protect{\ref{LHE}} (uniform BS configuration). During the $x$ evolution, the trapped area remains convex (in each  subinterval: $[(2n-1)\pi,(2n+1)\pi]$, $n \in \mathbb{Z}$) unlike the plots of fig.  \protect{\ref{Path1}}.} 
\end{figure}

On the other hand, as $x$ decreases, the homogenous solution $\phi_h^x$ which we explicitly index by $x$, traces a path in the space of functions. It begins from $\phi_h^{x\gg1}=C^{x \gg1}=-2^{7/6} R x^{-1/3}$ (low energies) and ends to $\phi_h^{x\ll1}=C^{x\ll 1}=-2^{3/2} R x^{-1/2}$ (high energies) corresponding to the top and bottom plots of fig. \ref{LHE} respectively. It is instructive, to plot the curve $C$ resulting from the first boundary condition for $\phi_h^{x\ll1}=C^{x\ll 1}$ (see fig. \ref{Path2}) and the second boundary condition for $\phi_h^x=C^{x\gg1}$ (see (\ref{bdry}) and figs. \ref{rest} and \ref{Path1}) as $x$ varies. Although strictly speaking, 
this is correct only in the two extreme cases, we still may gain some intuition about the transition from BH to BS assuming is valid for all values of $x$ (energies).
In both, fig. \ref{Path1} and fig. \ref{Path2} the transition from BH$\rightarrow$BS is depicted and looks similar: the BH increases in size and merges with the nearby BHs and becomes a BS warping around the compact dimension. In both cases, it seems that there is a singularity on the topology of the trapped horizon which appears as a cusp (see \cite{Kol:2002xz}) at $\theta=\pm \pi$. Certainly, the information we have here is incomplete. It is likely that the truth should lay somewhere between these two possibilities. However, it is notable that in both cases, a cusp on the trapped horizon at the merging point seems to appear. 

Regarding the convexity or non-convexity of the trapped surface fig. \ref{Path2} shows convexity for the intermediate energies and seems to agree with the current intuition in the literature (see for instance fig. 6 in \cite{Kol:2002xz}) about the way the transition to the BS takes place.  Fig. \ref{Path1} on the other hand, shows non-convexity. In fact, after the BS stage, there seem to appear holes inside the surface (see shaded regions in fig. \ref{Path1}); most likely due to the image charges (see section \ref{ES}). These holes are external to the trapped surface. We argue in the next section that most likely the transition has the (non-convex with holes) form of fig. \ref{Path1}.

\section{Towards the trapped boundary for all energies: finding $\phi_h$} \label{ES}
The complete solution, if exists, should respect the $r$, $\theta \leftrightarrow -\theta$ and  $\theta \leftrightarrow 2\pi+\theta$ symmetries. Thus, the only possible solution for $\nabla_{\perp}^2\phi_h=0$ is $\phi_h=$ constant which, according to the discussion of the previous section, does not work. Hence, we are forced to relax the condition $\nabla_{\perp}^2\phi_h=0$ to $\nabla_{\perp}^2\phi_h=$ ``image charge", provided that the ``charge(s)", lies outside the trapped surface and (also) respects the symmetries. In this case, the only possible solution consistent with the symmetries of the problem
has the form
\begin{align}\label{exac}
&\psi=\phi +C_x+\frac{1}{x u}\sum_i f_i \notag\\&
f_i \equiv \left[\frac{C_i(x)\sinh(u)}{\cosh(u)-\cos(\theta-\theta_i(x))}+(\theta \leftrightarrow -\theta) \right], \hspace{0.1in}u=r/R
\end{align}
where all the $\theta_i$'s should lay outside $C$. Two additional conditions are: (a) $\lim_{x \to \infty}\left[ C_x+2/x \sum_i f_i/(1-\cos(\theta_i(x))\right] \rightarrow C^{x\gg1}$ (low energies; small trapped surface) and (b) $\lim_{u, \to \infty ,x \to 0}\left[ C_x+1/(xu) \sum_i f_i\right] \rightarrow C^{x\ll1}$ (high energies; large trapped surface). A few remarks about the possible form of the trapped surface follow. 

\begin{itemize}

\item{Let is consider the range $\theta$ 
 in $[-\pi,\pi]$ and take into account that the image charges are located at $r=0$ and $\theta=\pm \theta_i$.  As $x$ decreases, the image charges should move to the right (left) if at a given $x$, the charges are already located right (left) of the $r$ axis. In the extreme limit where $x\rightarrow 0$, the sources should move at $\pm \pi$; that is at $\theta_i \rightarrow \pi$. For instance, in fig. \ref{rest}, the image charges at positive angles should lay (approximately) in the interval $\theta \in (2.25,\pi)$.}
\item{ 
Conditions (a) and (b) reproduce the low and the high energy limit behavior.}
\item{Since $C^{x\gg1}<0$ and $C^{x\ll1}<0$, conditions (a) and (b) imply that a subset of the image charges $C_i(x)$ (see (\ref{exac})) is possibly negative (a similar situation appears in \cite{Eardley:2002re}); i.e. the images correspond to negative energies. From an electrostatic analogy point of view, this is not a surprise as typically the image charges usually appear with opposite sign.}
\item{Previous point, implies that at small $r$ there will be a repulsive force between the image charges and the actual charge (located left and right of say $\theta=0$) causing the non-convexity of the trapped surface (see fig. \ref{Path1}). At the same time, the image charges located left and right of (say) $\theta=\pi$, will attract each other (being both negative!). These  would create an  
(external to the trapped surface,) surface. This surface of the image charges isolates them from the the trapped surface once  (see top plot of fig. \ref{Path1}) the trapped surface reaches the whole compact dimension (see two bottom plots of fig. \ref{Path1} and in particular the shaded areas). The surface that shields the image charges, is centered at $\theta=\pm \pi$ and shrinks to zero  (see last plot of fig. \ref{Path1}) as the energy tends to infinity.

 In other words, according to this scenario, there seem to exist holes inside the Black Holes.}
\end{itemize}

Concluding, we have argued that the topology should qualitatively change as in fig. \ref{Path1} contrary to the current intuition which is more compatible with fig. \ref{Path2}. The trapped boundary is certainly continuous but it appears a kink (cusp) at $\theta=\pi,r \ne 0$ and at the transition energy, it appears as not convex. The non-convexity is due to the repulsive action of the image charges. Certainly, a more thorough investigation is required to confirm or not our current intuition.
\section{Conclusions}
\begin{itemize}

\item{In this work, we study the evolution of the topology of the black object that will be formed during a shock-wave collision in the presence of extra dimensions following the Penrose method of trapped surfaces. It is emphasized that this method provides a lower bound on the extend of the actual horizon (apparent horizon). This implies that the black objects that we have studied are at least as large as the Penrose method predicts. Consequently, our conclusions (see below) might apply for lower but certainly not higher energies.} 

\item{ We find evidence that there will be a transition from the BH to the BS configuration as the energy increases. The transition occurs at a finite energy while the topology, based on our preliminary investigation, seems to exhibit a (cusp) singularity at the transition point. A mechanism of this transition is being proposed (see fig. \ref{Path1}). In particular, we argue about the possibility that the non-uniform BS\footnote{Non-uniform BS implies that the BS has energy greater but comparable with the one of the merging point.} will contain a cavity around $\theta=\pi$; that is as far as possible from the position of the (localized) energy of the colliding shocks (at $r=\theta=0$). This cavity surrounds the image charges and shrinks to zero as the energy increases; that is as the non-uniform BS tends to become a uniform BS.}

\item{The entropy\footnote{And possibly all the rest thermodynamical quantities.} of the black object (BS) at high enough energies behaves as if the extra dimensions are absent. In other words the entropy depends only on the extended dimensions in this limit. This agrees with the expectations of \cite{Giddings:2001bu}\footnote{We would like to thank S. Giddings for his correspondence and for pointing out this particular issue.}.}

\item{A BH is being created at lower energies while a BS is created at higher energies while there exists a merger point at some critical value of the energy corresponding to some $x$, called $x_m$. According to the two top (coinciding) curves of the lower panel of fig. \ref{merge}, $x_m>0.02$. According to(\ref{x})) $x_m>0.02$ implies that $16G_6E_m/L^3 \lesssim 29$ is a lower bound for the merging energy (see (\ref{etr})). Similarly, using the upper panel of fig. \ref{merge}, one observes that $x_h<8$ is a satisfactory bound where our analysis is (approximately) correct. This provides the higher bound on the energy that corresponds to the BH configuration. For $x=x_h=8$, equation (\ref{x}) implies $16G_6E_m/L^3\gtrsim 0.8$. These two (crude) bounds are the ones appearing in equation (\ref{etr}). This result is consistent with \cite{Sorkin:2003ka,Kudoh:2003ki,Kudoh:2004hs,Niarchos:2008jc,Emparan:2007wm,Harmark:2007md} .}

\item{On the other hand, according to the Gregory-Laflamme (GL) analysis  \cite{Gregory:1993vy,Gregory:1994bj,Gubser:2001ac} there is  an energy $E_{GL}$ where for lower energies, the {\it uniform} BS becomes unstable (GL instability). In five extended plus one compact dimensions, the corresponding (defined analogously to (\ref{etr})) dimensionless parameter $\mu_{GL}$, results to an unstable {\it uniform} BS when $\mu_{GL}\lesssim 3$.  We argue that in physical processes\footnote{Real life does not involve five extended dimensions but the argument still applies.} this instability might not appear: a black object (BH or BS) will be generally formed through some scattering. Now, according to the present investigation, a {\it uniform} BS is being formed for energies that saturate the upper bound of (\ref{etr}) or even higher ($\mu>29$) . For smaller energies, either a non-uniform BS or a BH is being formed. But this implies that there is no meaning in perturbing a {\it uniform} BS at energies (corresponding to) $\mu_{GL}=3$ or lower as this geometry can never be created (through a scattering) at this low energy; the perturbation of a  {\it uniform} BS at $\mu_{GL}=3$ thus seems to be meaningless \footnote{Formally, one may write a uniform BS solution for any small energy but it seems that dynamically, a low energy (almost) uniform BS can not be created.} in this set up.}

\item{ The BS configuration obviously is larger and hence  $S_{BS}>S_{BH}$. There is no meaning to compare the two for fixed energies $E$ as they are created and exist for different collision energy $2E$.}
\item{Figures \ref{Path1} and \ref{Path2} suggest that there exist an intermediate region. This corresponds to a non-uniform BS creation. Once the energy is increased further, it becomes uniform. Hence, the entropy inequalities become $S_{BS}>S_{non\hspace{0.01in}un.}>S_{BH}$.}

\item{Maybe a (Un. BS)$\rightarrow$(Non-Un. BS)$\rightarrow$(BH) transition is possible after the BS radiates enough energy but not the other way around. This does not contradict the maximum entropy principle as the total entropy, taking into account the entropy carried by the radiation, should increase in a BS$\rightarrow$BH process.}

\item{If extra dimensions are present, then the produced entropy bound $S_{trap}$ will generally be different from $S_{trap} \sim (\sqrt{s})^2$ ($\sqrt{s}$ is the c.m. energy; see (\ref{GE})) which applies for the real world, that is for $\mathbb{R}^{1,3}$ space-time. Equation $S_{trap} \sim (\sqrt{s})^2$ applies only in the extreme (high energy) limit where $G_D E/(R_i)^{D-3}\gg 1$ for all $i=1,2,...D-4$ and hence all the extra dimensions decouple. For lower energies, more extra dimensions will contribute to the energy dependence of $S_{trap}$ and hence to the produced entropy $S_{prod}$\footnote{That is $D$ in (\ref{GE}) equals $4+n$ where $n$ the number of compact dimensions that satisfy $G_D E/(R_i)^{D-3}\ll 1;i=1,...,n$.}. 
Hence, if the final entropy in a collision as a function of $\sqrt{s}$ is estimated, it could yield to information about the presence of extra dimensions \cite{Bleicher:2010qr}. The same reasoning applies to the AdS/CFT calculations which estimate total multiplicities at the LHC \cite{Kiritsis:2011yn,Lin:2010cb,Kovchegov:2009du,Gubser:2009sx,Gubser:2008pc}. Taking into account that the full string-theory is 10-dimensional as there exists a compact 5-dimensional compact manifold that surrounds the $AdS_5$ space, implies a possible change in the results of these works depending on the energy range of interest.}

\item{It would be interesting to find the full $\phi_h^x$ even numerically and trace the trajectory of the topology as the energy changes. In fact, the authors of \cite{DuenasVidal:2012sa,DuenasVidal:2010vi} which apply the numerical methods devised in \cite{Yoshino:2002tx}, have already developed techniques that may solve the boundary value problem of finding the trapped surface. Then, with the full (trapped) solution at hand one could investigate the proposals of this work and also search for any possible discontinuities in the thermodynamical quantities. In particular one could search for a third order phase transition in the entropy analogously to \cite{Kol:2002xz,Gross:1980he,Susskind:1997dr}. We leave this for a future investigation.}

\end{itemize}

\acknowledgments
We would like to thank Y. Constantinou, S. Giddings, E. Kiritsis, B. Kol, R. Meyer, K. Sfetsos, N. Toumbas and especially V. Niarchos for informative discussions. We also would like to thank The Un. of Cyprus, LPTENS and The Un. of Arizona for their warm hospitality during several stages of this effort.

This work was partially supported by European Union grants FP7-REGPOT-2008-1-CreteHEP
Cosmo-228644, and PERG07-GA-2010-268246 as well as EU program``Thalis" ESF/NSRF 2007-2013 .



\bibliography{references}              

\begin{thebibliography}{79}
\expandafter\ifx\csname natexlab\endcsname\relax\def\natexlab#1{#1}\fi
\expandafter\ifx\csname bibnamefont\endcsname\relax
  \def\bibnamefont#1{#1}\fi
\expandafter\ifx\csname bibfnamefont\endcsname\relax
  \def\bibfnamefont#1{#1}\fi
\expandafter\ifx\csname citenamefont\endcsname\relax
  \def\citenamefont#1{#1}\fi
\expandafter\ifx\csname url\endcsname\relax
  \def\url#1{\texttt{#1}}\fi
\expandafter\ifx\csname urlprefix\endcsname\relax\def\urlprefix{URL }\fi
\providecommand{\bibinfo}[2]{#2}
\providecommand{\eprint}[2][]{\url{#2}}

\bibitem[{\citenamefont{Myers and Perry}(1986)}]{Myers:1986un}
\bibinfo{author}{\bibfnamefont{R.~C.} \bibnamefont{Myers}} \bibnamefont{and}
  \bibinfo{author}{\bibfnamefont{M.}~\bibnamefont{Perry}},
  \bibinfo{journal}{Annals Phys.} \textbf{\bibinfo{volume}{172}},
  \bibinfo{pages}{304} (\bibinfo{year}{1986}).

\bibitem[{\citenamefont{Myers}(2011)}]{Myers:2011yc}
\bibinfo{author}{\bibfnamefont{R.~C.} \bibnamefont{Myers}}
  (\bibinfo{year}{2011}), \eprint{1111.1903}.

\bibitem[{\citenamefont{Emparan and Reall}(2002)}]{Emparan:2001wn}
\bibinfo{author}{\bibfnamefont{R.}~\bibnamefont{Emparan}} \bibnamefont{and}
  \bibinfo{author}{\bibfnamefont{H.~S.} \bibnamefont{Reall}},
  \bibinfo{journal}{Phys.Rev.Lett.} \textbf{\bibinfo{volume}{88}},
  \bibinfo{pages}{101101} (\bibinfo{year}{2002}), \eprint{hep-th/0110260}.

\bibitem[{\citenamefont{Kol}(2005)}]{Kol:2002xz}
\bibinfo{author}{\bibfnamefont{B.}~\bibnamefont{Kol}}, \bibinfo{journal}{JHEP}
  \textbf{\bibinfo{volume}{0510}}, \bibinfo{pages}{049} (\bibinfo{year}{2005}),
  \eprint{hep-th/0206220}.

\bibitem[{\citenamefont{Kol}(2006)}]{Kol:2004ww}
\bibinfo{author}{\bibfnamefont{B.}~\bibnamefont{Kol}},
  \bibinfo{journal}{Phys.Rept.} \textbf{\bibinfo{volume}{422}},
  \bibinfo{pages}{119} (\bibinfo{year}{2006}), \eprint{hep-th/0411240}.

\bibitem[{\citenamefont{Asnin et~al.}(2006)\citenamefont{Asnin, Kol, and
  Smolkin}}]{Asnin:2006ip}
\bibinfo{author}{\bibfnamefont{V.}~\bibnamefont{Asnin}},
  \bibinfo{author}{\bibfnamefont{B.}~\bibnamefont{Kol}}, \bibnamefont{and}
  \bibinfo{author}{\bibfnamefont{M.}~\bibnamefont{Smolkin}},
  \bibinfo{journal}{Class.Quant.Grav.} \textbf{\bibinfo{volume}{23}},
  \bibinfo{pages}{6805} (\bibinfo{year}{2006}), \eprint{hep-th/0607129}.

\bibitem[{\citenamefont{Kol and Wiseman}(2003)}]{Kol:2003ja}
\bibinfo{author}{\bibfnamefont{B.}~\bibnamefont{Kol}} \bibnamefont{and}
  \bibinfo{author}{\bibfnamefont{T.}~\bibnamefont{Wiseman}},
  \bibinfo{journal}{Class.Quant.Grav.} \textbf{\bibinfo{volume}{20}},
  \bibinfo{pages}{3493} (\bibinfo{year}{2003}), \eprint{hep-th/0304070}.

\bibitem[{\citenamefont{Kol et~al.}(2004)\citenamefont{Kol, Sorkin, and
  Piran}}]{Kol:2003if}
\bibinfo{author}{\bibfnamefont{B.}~\bibnamefont{Kol}},
  \bibinfo{author}{\bibfnamefont{E.}~\bibnamefont{Sorkin}}, \bibnamefont{and}
  \bibinfo{author}{\bibfnamefont{T.}~\bibnamefont{Piran}},
  \bibinfo{journal}{Phys.Rev.} \textbf{\bibinfo{volume}{D69}},
  \bibinfo{pages}{064031} (\bibinfo{year}{2004}), \eprint{hep-th/0309190}.

\bibitem[{\citenamefont{Gorbonos and Kol}(2005)}]{Gorbonos:2005px}
\bibinfo{author}{\bibfnamefont{D.}~\bibnamefont{Gorbonos}} \bibnamefont{and}
  \bibinfo{author}{\bibfnamefont{B.}~\bibnamefont{Kol}},
  \bibinfo{journal}{Class.Quant.Grav.} \textbf{\bibinfo{volume}{22}},
  \bibinfo{pages}{3935} (\bibinfo{year}{2005}), \eprint{hep-th/0505009}.

\bibitem[{\citenamefont{Emparan et~al.}(2007)\citenamefont{Emparan, Harmark,
  Niarchos, Obers, and Rodriguez}}]{Emparan:2007wm}
\bibinfo{author}{\bibfnamefont{R.}~\bibnamefont{Emparan}},
  \bibinfo{author}{\bibfnamefont{T.}~\bibnamefont{Harmark}},
  \bibinfo{author}{\bibfnamefont{V.}~\bibnamefont{Niarchos}},
  \bibinfo{author}{\bibfnamefont{N.~A.} \bibnamefont{Obers}}, \bibnamefont{and}
  \bibinfo{author}{\bibfnamefont{M.~J.} \bibnamefont{Rodriguez}},
  \bibinfo{journal}{JHEP} \textbf{\bibinfo{volume}{0710}}, \bibinfo{pages}{110}
  (\bibinfo{year}{2007}), \eprint{0708.2181}.

\bibitem[{\citenamefont{Emparan and Haddad}(2011)}]{Emparan:2011ve}
\bibinfo{author}{\bibfnamefont{R.}~\bibnamefont{Emparan}} \bibnamefont{and}
  \bibinfo{author}{\bibfnamefont{N.}~\bibnamefont{Haddad}},
  \bibinfo{journal}{JHEP} \textbf{\bibinfo{volume}{1110}}, \bibinfo{pages}{064}
  (\bibinfo{year}{2011}), \eprint{1109.1983}.

\bibitem[{\citenamefont{Eardley and Giddings}(2002)}]{Eardley:2002re}
\bibinfo{author}{\bibfnamefont{D.~M.} \bibnamefont{Eardley}} \bibnamefont{and}
  \bibinfo{author}{\bibfnamefont{S.~B.} \bibnamefont{Giddings}},
  \bibinfo{journal}{Phys. Rev.} \textbf{\bibinfo{volume}{D66}},
  \bibinfo{pages}{044011} (\bibinfo{year}{2002}), \eprint{gr-qc/0201034}.

\bibitem[{\citenamefont{Giddings and Rychkov}(2004)}]{Giddings:2004xy}
\bibinfo{author}{\bibfnamefont{S.~B.} \bibnamefont{Giddings}} \bibnamefont{and}
  \bibinfo{author}{\bibfnamefont{V.~S.} \bibnamefont{Rychkov}},
  \bibinfo{journal}{Phys. Rev.} \textbf{\bibinfo{volume}{D70}},
  \bibinfo{pages}{104026} (\bibinfo{year}{2004}), \eprint{hep-th/0409131}.

\bibitem[{\citenamefont{Kiritsis and Taliotis}(2011)}]{Kiritsis:2011yn}
\bibinfo{author}{\bibfnamefont{E.}~\bibnamefont{Kiritsis}} \bibnamefont{and}
  \bibinfo{author}{\bibfnamefont{A.}~\bibnamefont{Taliotis}}
  (\bibinfo{year}{2011}), \eprint{1111.1931}.

\bibitem[{\citenamefont{Aref'eva}(2009)}]{Aref'eva:2009ng}
\bibinfo{author}{\bibfnamefont{I.~Y.} \bibnamefont{Aref'eva}},
  \bibinfo{journal}{Theor. Math. Phys.} \textbf{\bibinfo{volume}{161}},
  \bibinfo{pages}{1647} (\bibinfo{year}{2009}), \eprint{0912.5481}.

\bibitem[{\citenamefont{Aref'eva et~al.}(2009)\citenamefont{Aref'eva, Bagrov,
  and Guseva}}]{Aref'eva:2009wz}
\bibinfo{author}{\bibfnamefont{I.~Y.} \bibnamefont{Aref'eva}},
  \bibinfo{author}{\bibfnamefont{A.~A.} \bibnamefont{Bagrov}},
  \bibnamefont{and} \bibinfo{author}{\bibfnamefont{E.~A.}
  \bibnamefont{Guseva}}, \bibinfo{journal}{JHEP} \textbf{\bibinfo{volume}{12}},
  \bibinfo{pages}{009} (\bibinfo{year}{2009}), \eprint{0905.1087}.

\bibitem[{\citenamefont{Aref'eva et~al.}(2010)\citenamefont{Aref'eva, Bagrov,
  and Joukovskaya}}]{Aref'eva:2009kw}
\bibinfo{author}{\bibfnamefont{I.~Y.} \bibnamefont{Aref'eva}},
  \bibinfo{author}{\bibfnamefont{A.~A.} \bibnamefont{Bagrov}},
  \bibnamefont{and} \bibinfo{author}{\bibfnamefont{L.~V.}
  \bibnamefont{Joukovskaya}}, \bibinfo{journal}{JHEP}
  \textbf{\bibinfo{volume}{03}}, \bibinfo{pages}{002} (\bibinfo{year}{2010}),
  \eprint{0909.1294}.

\bibitem[{\citenamefont{Nastase}(2008)}]{Nastase:2008hw}
\bibinfo{author}{\bibfnamefont{H.}~\bibnamefont{Nastase}},
  \bibinfo{journal}{Prog. Theor. Phys. Suppl.} \textbf{\bibinfo{volume}{174}},
  \bibinfo{pages}{274} (\bibinfo{year}{2008}), \eprint{0805.3579}.

\bibitem[{\citenamefont{Lin and Shuryak}(2010)}]{Lin:2010cb}
\bibinfo{author}{\bibfnamefont{S.}~\bibnamefont{Lin}} \bibnamefont{and}
  \bibinfo{author}{\bibfnamefont{E.}~\bibnamefont{Shuryak}}
  (\bibinfo{year}{2010}), \eprint{1011.1918}.

\bibitem[{\citenamefont{Kovchegov and Lin}(2010)}]{Kovchegov:2009du}
\bibinfo{author}{\bibfnamefont{Y.~V.} \bibnamefont{Kovchegov}}
  \bibnamefont{and} \bibinfo{author}{\bibfnamefont{S.}~\bibnamefont{Lin}},
  \bibinfo{journal}{JHEP} \textbf{\bibinfo{volume}{03}}, \bibinfo{pages}{057}
  (\bibinfo{year}{2010}), \eprint{0911.4707}.

\bibitem[{\citenamefont{Gubser et~al.}(2009)\citenamefont{Gubser, Pufu, and
  Yarom}}]{Gubser:2009sx}
\bibinfo{author}{\bibfnamefont{S.~S.} \bibnamefont{Gubser}},
  \bibinfo{author}{\bibfnamefont{S.~S.} \bibnamefont{Pufu}}, \bibnamefont{and}
  \bibinfo{author}{\bibfnamefont{A.}~\bibnamefont{Yarom}},
  \bibinfo{journal}{JHEP} \textbf{\bibinfo{volume}{11}}, \bibinfo{pages}{050}
  (\bibinfo{year}{2009}), \eprint{0902.4062}.

\bibitem[{\citenamefont{Gubser et~al.}(2008)\citenamefont{Gubser, Pufu, and
  Yarom}}]{Gubser:2008pc}
\bibinfo{author}{\bibfnamefont{S.~S.} \bibnamefont{Gubser}},
  \bibinfo{author}{\bibfnamefont{S.~S.} \bibnamefont{Pufu}}, \bibnamefont{and}
  \bibinfo{author}{\bibfnamefont{A.}~\bibnamefont{Yarom}},
  \bibinfo{journal}{Phys. Rev.} \textbf{\bibinfo{volume}{D78}},
  \bibinfo{pages}{066014} (\bibinfo{year}{2008}), \eprint{0805.1551}.

\bibitem[{\citenamefont{Duenas-Vidal and
  Vazquez-Mozo}(2012)}]{DuenasVidal:2012sa}
\bibinfo{author}{\bibfnamefont{A.}~\bibnamefont{Duenas-Vidal}}
  \bibnamefont{and} \bibinfo{author}{\bibfnamefont{M.~A.}
  \bibnamefont{Vazquez-Mozo}} (\bibinfo{year}{2012}), \eprint{1203.1046}.

\bibitem[{\citenamefont{Duenas-Vidal and
  Vazquez-Mozo}(2010)}]{DuenasVidal:2010vi}
\bibinfo{author}{\bibfnamefont{A.}~\bibnamefont{Duenas-Vidal}}
  \bibnamefont{and} \bibinfo{author}{\bibfnamefont{M.~A.}
  \bibnamefont{Vazquez-Mozo}}, \bibinfo{journal}{JHEP}
  \textbf{\bibinfo{volume}{1007}}, \bibinfo{pages}{021} (\bibinfo{year}{2010}),
  \eprint{1004.2609}.

\bibitem[{\citenamefont{Alvarez-Gaume et~al.}(2009)\citenamefont{Alvarez-Gaume,
  Gomez, Sabio~Vera, Tavanfar, and Vazquez-Mozo}}]{AlvarezGaume:2008fx}
\bibinfo{author}{\bibfnamefont{L.}~\bibnamefont{Alvarez-Gaume}},
  \bibinfo{author}{\bibfnamefont{C.}~\bibnamefont{Gomez}},
  \bibinfo{author}{\bibfnamefont{A.}~\bibnamefont{Sabio~Vera}},
  \bibinfo{author}{\bibfnamefont{A.}~\bibnamefont{Tavanfar}}, \bibnamefont{and}
  \bibinfo{author}{\bibfnamefont{M.~A.} \bibnamefont{Vazquez-Mozo}},
  \bibinfo{journal}{JHEP} \textbf{\bibinfo{volume}{02}}, \bibinfo{pages}{009}
  (\bibinfo{year}{2009}), \eprint{0811.3969}.

\bibitem[{\citenamefont{D'Eath and Payne}(1992{\natexlab{a}})}]{D'Eath:1992hb}
\bibinfo{author}{\bibfnamefont{P.~D.} \bibnamefont{D'Eath}} \bibnamefont{and}
  \bibinfo{author}{\bibfnamefont{P.~N.} \bibnamefont{Payne}},
  \bibinfo{journal}{Phys. Rev.} \textbf{\bibinfo{volume}{D46}},
  \bibinfo{pages}{658} (\bibinfo{year}{1992}{\natexlab{a}}).

\bibitem[{\citenamefont{D'Eath and Payne}(1992{\natexlab{b}})}]{D'Eath:1992hd}
\bibinfo{author}{\bibfnamefont{P.~D.} \bibnamefont{D'Eath}} \bibnamefont{and}
  \bibinfo{author}{\bibfnamefont{P.~N.} \bibnamefont{Payne}},
  \bibinfo{journal}{Phys. Rev.} \textbf{\bibinfo{volume}{D46}},
  \bibinfo{pages}{675} (\bibinfo{year}{1992}{\natexlab{b}}).

\bibitem[{\citenamefont{D'Eath and Payne}(1992{\natexlab{c}})}]{D'Eath:1992qu}
\bibinfo{author}{\bibfnamefont{P.~D.} \bibnamefont{D'Eath}} \bibnamefont{and}
  \bibinfo{author}{\bibfnamefont{P.~N.} \bibnamefont{Payne}},
  \bibinfo{journal}{Phys. Rev.} \textbf{\bibinfo{volume}{D46}},
  \bibinfo{pages}{694} (\bibinfo{year}{1992}{\natexlab{c}}).

\bibitem[{\citenamefont{Aichelburg and Sexl}(1971)}]{Aichelburg:1970dh}
\bibinfo{author}{\bibfnamefont{P.~C.} \bibnamefont{Aichelburg}}
  \bibnamefont{and} \bibinfo{author}{\bibfnamefont{R.~U.} \bibnamefont{Sexl}},
  \bibinfo{journal}{Gen. Rel. Grav.} \textbf{\bibinfo{volume}{2}},
  \bibinfo{pages}{303} (\bibinfo{year}{1971}).

\bibitem[{\citenamefont{Dray and 't~Hooft}(1985)}]{Dray:1984ha}
\bibinfo{author}{\bibfnamefont{T.}~\bibnamefont{Dray}} \bibnamefont{and}
  \bibinfo{author}{\bibfnamefont{G.}~\bibnamefont{'t~Hooft}},
  \bibinfo{journal}{Nucl. Phys.} \textbf{\bibinfo{volume}{B253}},
  \bibinfo{pages}{173} (\bibinfo{year}{1985}).

\bibitem[{\citenamefont{Hawking and Penrose}(1970)}]{Hawking:1969sw}
\bibinfo{author}{\bibfnamefont{S.~W.} \bibnamefont{Hawking}} \bibnamefont{and}
  \bibinfo{author}{\bibfnamefont{R.}~\bibnamefont{Penrose}},
  \bibinfo{journal}{Proc. Roy. Soc. Lond.} \textbf{\bibinfo{volume}{A314}},
  \bibinfo{pages}{529} (\bibinfo{year}{1970}).

\bibitem[{\citenamefont{Sfetsos}(1995)}]{Sfetsos:1994xa}
\bibinfo{author}{\bibfnamefont{K.}~\bibnamefont{Sfetsos}},
  \bibinfo{journal}{Nucl. Phys.} \textbf{\bibinfo{volume}{B436}},
  \bibinfo{pages}{721} (\bibinfo{year}{1995}), \eprint{hep-th/9408169}.

\bibitem[{\citenamefont{Taliotis}(2010{\natexlab{a}})}]{Taliotis:2010az}
\bibinfo{author}{\bibfnamefont{A.}~\bibnamefont{Taliotis}}
  (\bibinfo{year}{2010}{\natexlab{a}}), \eprint{1007.1452}.

\bibitem[{\citenamefont{Constantinou et~al.}(2011)\citenamefont{Constantinou,
  Gal'tsov, Spirin, and Tomaras}}]{Constantinou:2011ju}
\bibinfo{author}{\bibfnamefont{Y.}~\bibnamefont{Constantinou}},
  \bibinfo{author}{\bibfnamefont{D.}~\bibnamefont{Gal'tsov}},
  \bibinfo{author}{\bibfnamefont{P.}~\bibnamefont{Spirin}}, \bibnamefont{and}
  \bibinfo{author}{\bibfnamefont{T.~N.} \bibnamefont{Tomaras}},
  \bibinfo{journal}{JHEP} \textbf{\bibinfo{volume}{1111}}, \bibinfo{pages}{118}
  (\bibinfo{year}{2011}), \bibinfo{note}{26 pages, 7 figures},
  \eprint{1106.3509}.

\bibitem[{\citenamefont{Galtsov et~al.}(2010)\citenamefont{Galtsov, Kofinas,
  Spirin, and Tomaras}}]{Gal'tsov:2010me}
\bibinfo{author}{\bibfnamefont{D.~V.} \bibnamefont{Galtsov}},
  \bibinfo{author}{\bibfnamefont{G.}~\bibnamefont{Kofinas}},
  \bibinfo{author}{\bibfnamefont{P.}~\bibnamefont{Spirin}}, \bibnamefont{and}
  \bibinfo{author}{\bibfnamefont{T.~N.} \bibnamefont{Tomaras}},
  \bibinfo{journal}{JHEP} \textbf{\bibinfo{volume}{1005}}, \bibinfo{pages}{055}
  (\bibinfo{year}{2010}), \eprint{1003.2982}.

\bibitem[{\citenamefont{Gal'tsov et~al.}(2010)\citenamefont{Gal'tsov, Kofinas,
  Spirin, and Tomaras}}]{Gal'tsov:2009zi}
\bibinfo{author}{\bibfnamefont{D.~V.} \bibnamefont{Gal'tsov}},
  \bibinfo{author}{\bibfnamefont{G.}~\bibnamefont{Kofinas}},
  \bibinfo{author}{\bibfnamefont{P.}~\bibnamefont{Spirin}}, \bibnamefont{and}
  \bibinfo{author}{\bibfnamefont{T.~N.} \bibnamefont{Tomaras}},
  \bibinfo{journal}{Phys.Lett.} \textbf{\bibinfo{volume}{B683}},
  \bibinfo{pages}{331} (\bibinfo{year}{2010}), \eprint{0908.0675}.

\bibitem[{\citenamefont{Herdeiro et~al.}(2011)\citenamefont{Herdeiro, Sampaio,
  and Rebelo}}]{Herdeiro:2011ck}
\bibinfo{author}{\bibfnamefont{C.}~\bibnamefont{Herdeiro}},
  \bibinfo{author}{\bibfnamefont{M.~O.} \bibnamefont{Sampaio}},
  \bibnamefont{and} \bibinfo{author}{\bibfnamefont{C.}~\bibnamefont{Rebelo}},
  \bibinfo{journal}{JHEP} \textbf{\bibinfo{volume}{1107}}, \bibinfo{pages}{121}
  (\bibinfo{year}{2011}), \eprint{1105.2298}.

\bibitem[{\citenamefont{Aref'eva et~al.}(2012)\citenamefont{Aref'eva, Bagrov,
  and Pozdeeva}}]{Aref'eva:2012ar}
\bibinfo{author}{\bibfnamefont{I.}~\bibnamefont{Aref'eva}},
  \bibinfo{author}{\bibfnamefont{A.}~\bibnamefont{Bagrov}}, \bibnamefont{and}
  \bibinfo{author}{\bibfnamefont{E.}~\bibnamefont{Pozdeeva}}
  (\bibinfo{year}{2012}), \eprint{1201.6542}.

\bibitem[{\citenamefont{Kang and Nastase}(2005)}]{Kang:2004jd}
\bibinfo{author}{\bibfnamefont{K.}~\bibnamefont{Kang}} \bibnamefont{and}
  \bibinfo{author}{\bibfnamefont{H.}~\bibnamefont{Nastase}},
  \bibinfo{journal}{Phys.Rev.} \textbf{\bibinfo{volume}{D72}},
  \bibinfo{pages}{106003} (\bibinfo{year}{2005}), \eprint{hep-th/0410173}.

\bibitem[{\citenamefont{Giddings}(2003)}]{Giddings:2002cd}
\bibinfo{author}{\bibfnamefont{S.~B.} \bibnamefont{Giddings}},
  \bibinfo{journal}{Phys.Rev.} \textbf{\bibinfo{volume}{D67}},
  \bibinfo{pages}{126001} (\bibinfo{year}{2003}), \eprint{hep-th/0203004}.

\bibitem[{\citenamefont{Kovchegov and Taliotis}(2007)}]{Kovchegov:2007pq}
\bibinfo{author}{\bibfnamefont{Y.~V.} \bibnamefont{Kovchegov}}
  \bibnamefont{and} \bibinfo{author}{\bibfnamefont{A.}~\bibnamefont{Taliotis}},
  \bibinfo{journal}{Phys. Rev.} \textbf{\bibinfo{volume}{C76}},
  \bibinfo{pages}{014905} (\bibinfo{year}{2007}), \eprint{0705.1234}.

\bibitem[{\citenamefont{Spillane et~al.}(2011)\citenamefont{Spillane, Stoffers,
  and Zahed}}]{Spillane:2011yf}
\bibinfo{author}{\bibfnamefont{M.}~\bibnamefont{Spillane}},
  \bibinfo{author}{\bibfnamefont{A.}~\bibnamefont{Stoffers}}, \bibnamefont{and}
  \bibinfo{author}{\bibfnamefont{I.}~\bibnamefont{Zahed}}
  (\bibinfo{year}{2011}), \eprint{1110.5069}.

\bibitem[{\citenamefont{Hotta and Tanaka}(1993)}]{Hotta:1992qy}
\bibinfo{author}{\bibfnamefont{M.}~\bibnamefont{Hotta}} \bibnamefont{and}
  \bibinfo{author}{\bibfnamefont{M.}~\bibnamefont{Tanaka}},
  \bibinfo{journal}{Class. Quant. Grav.} \textbf{\bibinfo{volume}{10}},
  \bibinfo{pages}{307} (\bibinfo{year}{1993}).

\bibitem[{\citenamefont{Albacete
  et~al.}(2008{\natexlab{a}})\citenamefont{Albacete, Kovchegov, and
  Taliotis}}]{Albacete:2008vs}
\bibinfo{author}{\bibfnamefont{J.~L.} \bibnamefont{Albacete}},
  \bibinfo{author}{\bibfnamefont{Y.~V.} \bibnamefont{Kovchegov}},
  \bibnamefont{and} \bibinfo{author}{\bibfnamefont{A.}~\bibnamefont{Taliotis}},
  \bibinfo{journal}{JHEP} \textbf{\bibinfo{volume}{07}}, \bibinfo{pages}{100}
  (\bibinfo{year}{2008}{\natexlab{a}}), \eprint{0805.2927}.

\bibitem[{\citenamefont{Khlebnikov et~al.}(2010)\citenamefont{Khlebnikov,
  Kruczenski, and Michalogiorgakis}}]{Khlebnikov:2010yt}
\bibinfo{author}{\bibfnamefont{S.}~\bibnamefont{Khlebnikov}},
  \bibinfo{author}{\bibfnamefont{M.}~\bibnamefont{Kruczenski}},
  \bibnamefont{and}
  \bibinfo{author}{\bibfnamefont{G.}~\bibnamefont{Michalogiorgakis}}
  (\bibinfo{year}{2010}), \eprint{1004.3803}.

\bibitem[{\citenamefont{Khlebnikov et~al.}(2011)\citenamefont{Khlebnikov,
  Kruczenski, and Michalogiorgakis}}]{Khlebnikov:2011ka}
\bibinfo{author}{\bibfnamefont{S.}~\bibnamefont{Khlebnikov}},
  \bibinfo{author}{\bibfnamefont{M.}~\bibnamefont{Kruczenski}},
  \bibnamefont{and}
  \bibinfo{author}{\bibfnamefont{G.}~\bibnamefont{Michalogiorgakis}},
  \bibinfo{journal}{JHEP} \textbf{\bibinfo{volume}{1107}}, \bibinfo{pages}{097}
  (\bibinfo{year}{2011}), \eprint{1105.1355}.

\bibitem[{\citenamefont{Albacete et~al.}(2009)\citenamefont{Albacete,
  Kovchegov, and Taliotis}}]{Albacete:2009ji}
\bibinfo{author}{\bibfnamefont{J.~L.} \bibnamefont{Albacete}},
  \bibinfo{author}{\bibfnamefont{Y.~V.} \bibnamefont{Kovchegov}},
  \bibnamefont{and} \bibinfo{author}{\bibfnamefont{A.}~\bibnamefont{Taliotis}},
  \bibinfo{journal}{JHEP} \textbf{\bibinfo{volume}{05}}, \bibinfo{pages}{060}
  (\bibinfo{year}{2009}), \eprint{0902.3046}.

\bibitem[{\citenamefont{Wu and Romatschke}(2011)}]{Wu:2011yd}
\bibinfo{author}{\bibfnamefont{B.}~\bibnamefont{Wu}} \bibnamefont{and}
  \bibinfo{author}{\bibfnamefont{P.}~\bibnamefont{Romatschke}},
  \bibinfo{journal}{Int.J.Mod.Phys.} \textbf{\bibinfo{volume}{C22}},
  \bibinfo{pages}{1317} (\bibinfo{year}{2011}), \eprint{1108.3715}.

\bibitem[{\citenamefont{Albacete
  et~al.}(2008{\natexlab{b}})\citenamefont{Albacete, Kovchegov, and
  Taliotis}}]{Albacete:2008ze}
\bibinfo{author}{\bibfnamefont{J.~L.} \bibnamefont{Albacete}},
  \bibinfo{author}{\bibfnamefont{Y.~V.} \bibnamefont{Kovchegov}},
  \bibnamefont{and} \bibinfo{author}{\bibfnamefont{A.}~\bibnamefont{Taliotis}},
  \bibinfo{journal}{JHEP} \textbf{\bibinfo{volume}{07}}, \bibinfo{pages}{074}
  (\bibinfo{year}{2008}{\natexlab{b}}), \eprint{0806.1484}.

\bibitem[{\citenamefont{Taliotis}(2009)}]{Taliotis:2009ne}
\bibinfo{author}{\bibfnamefont{A.}~\bibnamefont{Taliotis}},
  \bibinfo{journal}{Nucl. Phys.} \textbf{\bibinfo{volume}{A830}},
  \bibinfo{pages}{299c} (\bibinfo{year}{2009}), \eprint{0907.4204}.

\bibitem[{\citenamefont{Grumiller and Romatschke}(2008)}]{Grumiller:2008va}
\bibinfo{author}{\bibfnamefont{D.}~\bibnamefont{Grumiller}} \bibnamefont{and}
  \bibinfo{author}{\bibfnamefont{P.}~\bibnamefont{Romatschke}},
  \bibinfo{journal}{JHEP} \textbf{\bibinfo{volume}{08}}, \bibinfo{pages}{027}
  (\bibinfo{year}{2008}), \eprint{0803.3226}.

\bibitem[{\citenamefont{Taliotis}(2010{\natexlab{b}})}]{Taliotis:2010pi}
\bibinfo{author}{\bibfnamefont{A.}~\bibnamefont{Taliotis}},
  \bibinfo{journal}{JHEP} \textbf{\bibinfo{volume}{09}}, \bibinfo{pages}{102}
  (\bibinfo{year}{2010}{\natexlab{b}}), \eprint{1004.3500}.

\bibitem[{\citenamefont{Harmark and Obers}(2002)}]{Harmark:2002tr}
\bibinfo{author}{\bibfnamefont{T.}~\bibnamefont{Harmark}} \bibnamefont{and}
  \bibinfo{author}{\bibfnamefont{N.~A.} \bibnamefont{Obers}},
  \bibinfo{journal}{JHEP} \textbf{\bibinfo{volume}{0205}}, \bibinfo{pages}{032}
  (\bibinfo{year}{2002}), \eprint{hep-th/0204047}.

\bibitem[{\citenamefont{Harmark and Obers}(2004)}]{Harmark:2003eg}
\bibinfo{author}{\bibfnamefont{T.}~\bibnamefont{Harmark}} \bibnamefont{and}
  \bibinfo{author}{\bibfnamefont{N.~A.} \bibnamefont{Obers}},
  \bibinfo{journal}{Nucl.Phys.} \textbf{\bibinfo{volume}{B684}},
  \bibinfo{pages}{183} (\bibinfo{year}{2004}), \eprint{hep-th/0309230}.

\bibitem[{\citenamefont{Karasik et~al.}(2005)\citenamefont{Karasik, Sahabandu,
  Suranyi, and Wijewardhana}}]{Karasik:2004ds}
\bibinfo{author}{\bibfnamefont{D.}~\bibnamefont{Karasik}},
  \bibinfo{author}{\bibfnamefont{C.}~\bibnamefont{Sahabandu}},
  \bibinfo{author}{\bibfnamefont{P.}~\bibnamefont{Suranyi}}, \bibnamefont{and}
  \bibinfo{author}{\bibfnamefont{L.}~\bibnamefont{Wijewardhana}},
  \bibinfo{journal}{Phys.Rev.} \textbf{\bibinfo{volume}{D71}},
  \bibinfo{pages}{024024} (\bibinfo{year}{2005}), \bibinfo{note}{21 pages, 4
  figures. Replaces previous version, with added references and slightly
  altered discussion}, \eprint{hep-th/0410078}.

\bibitem[{\citenamefont{Chu et~al.}(2006)\citenamefont{Chu, Goldberger, and
  Rothstein}}]{Chu:2006ce}
\bibinfo{author}{\bibfnamefont{Y.-Z.} \bibnamefont{Chu}},
  \bibinfo{author}{\bibfnamefont{W.~D.} \bibnamefont{Goldberger}},
  \bibnamefont{and} \bibinfo{author}{\bibfnamefont{I.~Z.}
  \bibnamefont{Rothstein}}, \bibinfo{journal}{JHEP}
  \textbf{\bibinfo{volume}{0603}}, \bibinfo{pages}{013} (\bibinfo{year}{2006}),
  \eprint{hep-th/0602016}.

\bibitem[{\citenamefont{Wiseman}(2003)}]{Wiseman:2002zc}
\bibinfo{author}{\bibfnamefont{T.}~\bibnamefont{Wiseman}},
  \bibinfo{journal}{Class.Quant.Grav.} \textbf{\bibinfo{volume}{20}},
  \bibinfo{pages}{1137} (\bibinfo{year}{2003}), \eprint{hep-th/0209051}.

\bibitem[{\citenamefont{Sorkin}(2004)}]{Sorkin:2004qq}
\bibinfo{author}{\bibfnamefont{E.}~\bibnamefont{Sorkin}},
  \bibinfo{journal}{Phys.Rev.Lett.} \textbf{\bibinfo{volume}{93}},
  \bibinfo{pages}{031601} (\bibinfo{year}{2004}), \eprint{hep-th/0402216}.

\bibitem[{\citenamefont{Kleihaus et~al.}(2006)\citenamefont{Kleihaus, Kunz, and
  Radu}}]{Kleihaus:2006ee}
\bibinfo{author}{\bibfnamefont{B.}~\bibnamefont{Kleihaus}},
  \bibinfo{author}{\bibfnamefont{J.}~\bibnamefont{Kunz}}, \bibnamefont{and}
  \bibinfo{author}{\bibfnamefont{E.}~\bibnamefont{Radu}},
  \bibinfo{journal}{JHEP} \textbf{\bibinfo{volume}{0606}}, \bibinfo{pages}{016}
  (\bibinfo{year}{2006}), \eprint{hep-th/0603119}.

\bibitem[{\citenamefont{Sorkin et~al.}(2004)\citenamefont{Sorkin, Kol, and
  Piran}}]{Sorkin:2003ka}
\bibinfo{author}{\bibfnamefont{E.}~\bibnamefont{Sorkin}},
  \bibinfo{author}{\bibfnamefont{B.}~\bibnamefont{Kol}}, \bibnamefont{and}
  \bibinfo{author}{\bibfnamefont{T.}~\bibnamefont{Piran}},
  \bibinfo{journal}{Phys.Rev.} \textbf{\bibinfo{volume}{D69}},
  \bibinfo{pages}{064032} (\bibinfo{year}{2004}), \eprint{hep-th/0310096}.

\bibitem[{\citenamefont{Kudoh and Wiseman}(2004)}]{Kudoh:2003ki}
\bibinfo{author}{\bibfnamefont{H.}~\bibnamefont{Kudoh}} \bibnamefont{and}
  \bibinfo{author}{\bibfnamefont{T.}~\bibnamefont{Wiseman}},
  \bibinfo{journal}{Prog.Theor.Phys.} \textbf{\bibinfo{volume}{111}},
  \bibinfo{pages}{475} (\bibinfo{year}{2004}), \eprint{hep-th/0310104}.

\bibitem[{\citenamefont{Kudoh and Wiseman}(2005)}]{Kudoh:2004hs}
\bibinfo{author}{\bibfnamefont{H.}~\bibnamefont{Kudoh}} \bibnamefont{and}
  \bibinfo{author}{\bibfnamefont{T.}~\bibnamefont{Wiseman}},
  \bibinfo{journal}{Phys.Rev.Lett.} \textbf{\bibinfo{volume}{94}},
  \bibinfo{pages}{161102} (\bibinfo{year}{2005}), \eprint{hep-th/0409111}.

\bibitem[{\citenamefont{Harmark et~al.}(2007)\citenamefont{Harmark, Niarchos,
  and Obers}}]{Harmark:2007md}
\bibinfo{author}{\bibfnamefont{T.}~\bibnamefont{Harmark}},
  \bibinfo{author}{\bibfnamefont{V.}~\bibnamefont{Niarchos}}, \bibnamefont{and}
  \bibinfo{author}{\bibfnamefont{N.~A.} \bibnamefont{Obers}},
  \bibinfo{journal}{Class.Quant.Grav.} \textbf{\bibinfo{volume}{24}},
  \bibinfo{pages}{R1} (\bibinfo{year}{2007}), \eprint{hep-th/0701022}.

\bibitem[{\citenamefont{Chowdhury et~al.}(2007)\citenamefont{Chowdhury, Giusto,
  and Mathur}}]{Chowdhury:2006qn}
\bibinfo{author}{\bibfnamefont{B.~D.} \bibnamefont{Chowdhury}},
  \bibinfo{author}{\bibfnamefont{S.}~\bibnamefont{Giusto}}, \bibnamefont{and}
  \bibinfo{author}{\bibfnamefont{S.~D.} \bibnamefont{Mathur}},
  \bibinfo{journal}{Nucl.Phys.} \textbf{\bibinfo{volume}{B762}},
  \bibinfo{pages}{301} (\bibinfo{year}{2007}), \eprint{hep-th/0610069}.

\bibitem[{\citenamefont{Mathur}(2005)}]{Mathur:2005zp}
\bibinfo{author}{\bibfnamefont{S.~D.} \bibnamefont{Mathur}},
  \bibinfo{journal}{Fortsch.Phys.} \textbf{\bibinfo{volume}{53}},
  \bibinfo{pages}{793} (\bibinfo{year}{2005}), \eprint{hep-th/0502050}.

\bibitem[{\citenamefont{Mathur}(2006)}]{Mathur:2005ai}
\bibinfo{author}{\bibfnamefont{S.~D.} \bibnamefont{Mathur}},
  \bibinfo{journal}{Class.Quant.Grav.} \textbf{\bibinfo{volume}{23}},
  \bibinfo{pages}{R115} (\bibinfo{year}{2006}), \eprint{hep-th/0510180}.

\bibitem[{\citenamefont{Randall and Sundrum}(1999)}]{Randall:1999ee}
\bibinfo{author}{\bibfnamefont{L.}~\bibnamefont{Randall}} \bibnamefont{and}
  \bibinfo{author}{\bibfnamefont{R.}~\bibnamefont{Sundrum}},
  \bibinfo{journal}{Phys.Rev.Lett.} \textbf{\bibinfo{volume}{83}},
  \bibinfo{pages}{3370} (\bibinfo{year}{1999}), \bibinfo{note}{9 pages, LaTex
  Report-no: MIT-CTP-2860, PUPT-1860, BUHEP-99-9}, \eprint{hep-ph/9905221}.

\bibitem[{\citenamefont{Antoniadis et~al.}(1998)\citenamefont{Antoniadis,
  Arkani-Hamed, Dimopoulos, and Dvali}}]{Antoniadis:1998ig}
\bibinfo{author}{\bibfnamefont{I.}~\bibnamefont{Antoniadis}},
  \bibinfo{author}{\bibfnamefont{N.}~\bibnamefont{Arkani-Hamed}},
  \bibinfo{author}{\bibfnamefont{S.}~\bibnamefont{Dimopoulos}},
  \bibnamefont{and} \bibinfo{author}{\bibfnamefont{G.}~\bibnamefont{Dvali}},
  \bibinfo{journal}{Phys.Lett.} \textbf{\bibinfo{volume}{B436}},
  \bibinfo{pages}{257} (\bibinfo{year}{1998}), \eprint{hep-ph/9804398}.

\bibitem[{\citenamefont{Arkani-Hamed et~al.}(1998)\citenamefont{Arkani-Hamed,
  Dimopoulos, and Dvali}}]{ArkaniHamed:1998rs}
\bibinfo{author}{\bibfnamefont{N.}~\bibnamefont{Arkani-Hamed}},
  \bibinfo{author}{\bibfnamefont{S.}~\bibnamefont{Dimopoulos}},
  \bibnamefont{and} \bibinfo{author}{\bibfnamefont{G.}~\bibnamefont{Dvali}},
  \bibinfo{journal}{Phys.Lett.} \textbf{\bibinfo{volume}{B429}},
  \bibinfo{pages}{263} (\bibinfo{year}{1998}), \eprint{hep-ph/9803315}.

\bibitem[{\citenamefont{Arkani-Hamed et~al.}(1999)\citenamefont{Arkani-Hamed,
  Dimopoulos, and Dvali}}]{ArkaniHamed:1998nn}
\bibinfo{author}{\bibfnamefont{N.}~\bibnamefont{Arkani-Hamed}},
  \bibinfo{author}{\bibfnamefont{S.}~\bibnamefont{Dimopoulos}},
  \bibnamefont{and} \bibinfo{author}{\bibfnamefont{G.}~\bibnamefont{Dvali}},
  \bibinfo{journal}{Phys.Rev.} \textbf{\bibinfo{volume}{D59}},
  \bibinfo{pages}{086004} (\bibinfo{year}{1999}), \eprint{hep-ph/9807344}.

\bibitem[{\citenamefont{Giddings and Thomas}(2002)}]{Giddings:2001bu}
\bibinfo{author}{\bibfnamefont{S.~B.} \bibnamefont{Giddings}} \bibnamefont{and}
  \bibinfo{author}{\bibfnamefont{S.~D.} \bibnamefont{Thomas}},
  \bibinfo{journal}{Phys. Rev.} \textbf{\bibinfo{volume}{D65}},
  \bibinfo{pages}{056010} (\bibinfo{year}{2002}), \eprint{hep-ph/0106219}.

\bibitem[{\citenamefont{Niarchos}(2008)}]{Niarchos:2008jc}
\bibinfo{author}{\bibfnamefont{V.}~\bibnamefont{Niarchos}},
  \bibinfo{journal}{Mod.Phys.Lett.} \textbf{\bibinfo{volume}{A23}},
  \bibinfo{pages}{2625} (\bibinfo{year}{2008}), \eprint{0808.2776}.

\bibitem[{\citenamefont{Gregory and Laflamme}(1993)}]{Gregory:1993vy}
\bibinfo{author}{\bibfnamefont{R.}~\bibnamefont{Gregory}} \bibnamefont{and}
  \bibinfo{author}{\bibfnamefont{R.}~\bibnamefont{Laflamme}},
  \bibinfo{journal}{Phys.Rev.Lett.} \textbf{\bibinfo{volume}{70}},
  \bibinfo{pages}{2837} (\bibinfo{year}{1993}), \eprint{hep-th/9301052}.

\bibitem[{\citenamefont{Gregory and Laflamme}(1994)}]{Gregory:1994bj}
\bibinfo{author}{\bibfnamefont{R.}~\bibnamefont{Gregory}} \bibnamefont{and}
  \bibinfo{author}{\bibfnamefont{R.}~\bibnamefont{Laflamme}},
  \bibinfo{journal}{Nucl.Phys.} \textbf{\bibinfo{volume}{B428}},
  \bibinfo{pages}{399} (\bibinfo{year}{1994}), \eprint{hep-th/9404071}.

\bibitem[{\citenamefont{Gubser}(2002)}]{Gubser:2001ac}
\bibinfo{author}{\bibfnamefont{S.~S.} \bibnamefont{Gubser}},
  \bibinfo{journal}{Class.Quant.Grav.} \textbf{\bibinfo{volume}{19}},
  \bibinfo{pages}{4825} (\bibinfo{year}{2002}), \eprint{hep-th/0110193}.

\bibitem[{\citenamefont{Bleicher and Nicolini}(2010)}]{Bleicher:2010qr}
\bibinfo{author}{\bibfnamefont{M.}~\bibnamefont{Bleicher}} \bibnamefont{and}
  \bibinfo{author}{\bibfnamefont{P.}~\bibnamefont{Nicolini}},
  \bibinfo{journal}{J.Phys.Conf.Ser.} \textbf{\bibinfo{volume}{237}},
  \bibinfo{pages}{012008} (\bibinfo{year}{2010}), \eprint{1001.2211}.

\bibitem[{\citenamefont{Yoshino and Nambu}(2003)}]{Yoshino:2002tx}
\bibinfo{author}{\bibfnamefont{H.}~\bibnamefont{Yoshino}} \bibnamefont{and}
  \bibinfo{author}{\bibfnamefont{Y.}~\bibnamefont{Nambu}},
  \bibinfo{journal}{Phys.Rev.} \textbf{\bibinfo{volume}{D67}},
  \bibinfo{pages}{024009} (\bibinfo{year}{2003}), \eprint{gr-qc/0209003}.

\bibitem[{\citenamefont{Gross and Witten}(1980)}]{Gross:1980he}
\bibinfo{author}{\bibfnamefont{D.}~\bibnamefont{Gross}} \bibnamefont{and}
  \bibinfo{author}{\bibfnamefont{E.}~\bibnamefont{Witten}},
  \bibinfo{journal}{Phys.Rev.} \textbf{\bibinfo{volume}{D21}},
  \bibinfo{pages}{446} (\bibinfo{year}{1980}).

\bibitem[{\citenamefont{Susskind}(1997)}]{Susskind:1997dr}
\bibinfo{author}{\bibfnamefont{L.}~\bibnamefont{Susskind}}
  (\bibinfo{year}{1997}), \eprint{hep-th/9805115}.

\end{thebibliography}

\end{document}